\documentclass{aa}
\usepackage{hyperref}
\usepackage{txfonts}
\usepackage{graphicx}
\usepackage{float}
\usepackage{natbib}
\bibpunct{(}{)}{;}{a}{}{,}
\usepackage[english]{babel}

\title{Modeling the evolution of infrared galaxies: A Parametric backwards evolution model}

\author{M. B\'ethermin\inst{1,2} \and H. Dole\inst{1,2} \and G. Lagache\inst{1,2} \and D. Le Borgne \inst{3} \and A. Penin \inst{1,2}}

\institute{
\inst{1} Univ Paris-Sud, Laboratoire IAS, UMR8617, Orsay, F-91405\\
e-mail: matthieu.bethermin@ias.u-psud.fr\\
\inst{2} CNRS, Orsay, F-91405\\
\inst{3} Institut d'Astrophysique de Paris (IAP), UMR 7095 CNRS, UPMC, 98 bis boulevard Arago, F-75014 Paris, France }

\date{Submitted 30 September 2010 / Accepted 21 January 2011}

\abstract{}
{We aim at modeling the infrared galaxy evolution in a way as simple as possible and reproduce statistical properties among which the number counts between 15~$\mu$m and 1.1~mm, the luminosity functions, and the redshift distributions. We then aim at using this model to interpret the recent observations (\textit{Spitzer}, Akari, BLAST, LABOCA, AzTEC, SPT and \textit{Herschel}), and make predictions for \textit{Planck} and future experiments like CCAT or SPICA.
}{
This model uses an evolution in density and luminosity of the luminosity function parametrized by broken power-laws with two breaks at redshift $\sim$0.9 and 2, and contains the two populations of the Lagache et al. (2004) model: normal and starburst galaxies. We also take into account the effect of the strong lensing of high-redshift sub-millimeter galaxies. This effect is significant in the sub-mm and mm range near 50~mJy. It has 13 free parameters and 8 additional calibration parameters. We fit the parameters to the IRAS, \textit{Spitzer}, \textit{Herschel} and \textit{AzTEC} measurements with a Monte-Carlo Markov chain.
}{
The model ajusted on deep counts at key wavelengths reproduces the counts from the mid-infrared to the millimeter wavelengths, as well as the mid-infrared luminosity functions. We discuss the contribution to the cosmic infrared background (CIB) and to the infrared luminosity density of the different populations. We also estimate the effect of the lensing on the number counts, and discuss the recent discovery by the South Pole Telescope (SPT) of a very bright population lying at high-redshift. We predict the contribution of the lensed sources to the \textit{Planck} number counts, the confusion level for future missions using a P(D) formalism, and the Universe opacity to TeV photons due to the CIB. Material of the model (software, tables and predictions) is available at http://www.ias.u-psud.fr/irgalaxies/.
}{}

\titlerunning{Parametric backwards evolution model}
\authorrunning{B\'ethermin et al.}

\begin{document}

\maketitle

\section{Introduction}

The extragalactic background light (EBL) is the relic emission due to galaxy formation and accretion processes since the recombination. The infrared ($8~\mu m < \lambda < 1000~\mu m$) part of this emission called cosmic infrared background (CIB) was detected for the first time by \citet{Puget1996} and contains about half of the energy of the EBL \citep{Dole2006,Bethermin2010a}. Nevertheless, in the local universe, the optical/UV emissions are 3 times larger than infrared/sub-millimeter ones \citep{Soifer1991,Driver2008}. This pseudo-paradox is explained by a strong evolution of the properties of the infrared galaxies.\\

The infrared luminosity density is dominated by normal galaxies ($L_{IR,bolometric} < 10^{11} L_{\odot}$) in the local Universe \citep{Saunders1990}. At higher redshift, it is dominated by luminous infrared galaxies (LIRG, $10^{11} L_{\odot} < L_{IR,bolometric} < 10^{12} L_{\odot}$) at z=1 \citep{Le_Floch2005} and by ultra-luminous infrared galaxies (ULIRG, $10^{12} L_{\odot} < L_{IR,bolometric} < 10^{13} L_{\odot}$) at z=2 \citep{Caputi2007}. The infrared luminosity of these galaxies is correlated to the star formation rate \citep{Kennicutt1998}. Thus modeling this rapid evolution of the infrared galaxies is very important to understand the history of the star formation.\\

The physical models (such as \citet{Lacey2010,Wilman2010,Younger2010} for the latest) use a physical approach based on semi-analytical recipes and dark matter numerical simulations. They use a limited set of physical parameters, but they nowadays poorly reproduce some basic observational constraints like the infrared galaxy number counts \citep{Oliver2010}.\\

The backwards evolution models (like \citet{Lagache2004,Franceschini2009,Rowan2009,Valiante2009}) use an evolution the luminosity function (LF) of the galaxies to reproduce empirically the galaxy counts, and other constraints. These models make only a description of the evolution and contain little physics. The parameters of these models were tuned manually to fit observational constraints. \citet{Le_Borgne2009} used an other approach and performed a non-parametric inversion of the counts to determine the LF. Nevertheless, this approach is complex, uses only one population of galaxy, and does not manage to reproduce the 160~$\mu$m number counts. An other fully-empirical approach was used by \citet{Dominguez2010}. They fitted the SED from UV to mid-infrared of detected galaxies and extrapolated the far-infrared spectral energy distribution of these galaxies and the contribution of faint populations. Nevertheless, their model aims only to reproduce the CIB; however its ability to reproduce other constraints like the number counts was not tested.\\

The Balloon-borne Large-Aperture Submillimeter Telescope (BLAST) experiment \citep{Pascale2008,Devlin2009} and the spectral and photometric imaging receiver (SPIRE) instrument \citep{Griffin2010} onboard the \textit{Herchel} space telescope \citep{Pilbratt2010} performed recently new observations in the sub-mm at 250, 350 and 500~$\mu$m. In their current version, most of the models fail to reproduce the number counts measured at these wavelengths \citep{Patanchon2009,Bethermin2010b,Clements2010,Oliver2010}. The \citet{Valiante2009} model gives the best results, using a Monte Carlo approach (sources are randomly taken in libraries) to simulate the temperature scatter and the heterogeneity of the populations of active galactic nucleus (AGN), but this model strongly disagrees with the recent measurements of the redshift distribution of the CIB by \citet{Jauzac2010}. It is thus necessary to develop new models that reproduce the recent far-infrared and sub-mm observations.\\

The discovery of very bright and high-redshift dusty galaxies by \citet{Vieira2009} with the south pole telescope (SPT) suggests that the contribution of high-redshift galaxies strongly lensed by dark matter halos of massive low-redshift galaxies on the bright sub-millimeter and millimeter counts is non negligible. This contribution was discussed by \citet{Negrello2007} and an observational evidence of this phenomenon was found very recently by \citet{Negrello2010}. We can also cite the simplified approach of \citet{Lima2010} who reproduce the AzTEC and SPT counts using a single population of galaxies with a Schechter LF at a single redshift and a lensing model. We can also cite the very recent work of \citet{Hezaveh2010} on the effect of the lensing on the SPT counts, based on an advanced lensing model. \\

We present a new simple and parametric model based on \citet{Lagache2004} SED libraries, which reproduces the new observational constraints. The parameters of this model (13 free parameters and 8 calibration parameters) were fitted from a large set of recent observations using a Monte-Carlo Markov chain (MCMC) method, allowing to study degeneracies between the parameters. This model also includes the effects of the strong lensing on the observations. We make predictions on the confusion limit for future missions, on the high-energy opacity of the Universe and on the effects of the strong lensing on the counts. This model is plugged to a halo model to study the spatial distribution of the infrared galaxies in a companion paper \citep{Penin2010}. Note that an other study using also MCMC methods was performed by \citet{Marsden2010} at the same time than ours.\\

We use the WMAP 7 year best-fit $\Lambda$CDM cosmology in this paper \citep{Larson2010}. We thus have H$_0$ = 71~km.s$^{-1}$.Mpc$^{-1}$, $\Omega_\Lambda$ = 0.734 and $\Omega_m$ = 0.266.

\section{Approach}

The backward evolution models are not built on physical parameters. Each model uses different evolving populations to reproduce the observational constraints. Some recent models (like \citet{Franceschini2009,Rowan2009}) use 4 galaxy populations evolving separately to reproduce the observations. \citet{Valiante2009} take randomly galaxy SEDs on a very large library of templates and claim that the contribution of the AGNs and the dispersion of the dust temperature of the galaxies must be taken into account to reproduce the observational constraint. Our approach is to keep the model as simple as possible, but to use advanced methods to constrain its free parameters. This new parametric model can be used as an input for halo modeling or P(D) analysis for instance.\\

As it will be shown, we did not need AGN contribution and temperature dispersion to reproduce the current observational constraints. In fact, in the local Universe, the AGNs only dominate the ULIRG regime \citep{Imanishi2009}. \citet{Alexander2005} estimate an AGN contribution of 8\% for the submillimeter galaxies (SMG). Recently \citet{Fadda2010} showed that the proportion of AGN-dominated sources is rather small for LIRGs at z$\sim$1 (5\%) and ULIRGs around z$\sim$2 (12\%). \citet{Jauzac2010} showed that AGN contribution to the CIB is less than 10\% at z$<$1.5. These category of luminosity dominates the infrared output at their redshift. The low contribution of AGN in these categories explains why the AGNs are not necessary to reproduce the mean statistical properties of the galaxies. Nevertheless, despite their small contribution to the infrared output, the AGNs play a central role in the physics of galaxies.\\

Our model takes into account the the strong-lensing of high redshift galaxies by dark matter halos of elliptical galaxies. According to the results of Sect. \ref{section:stronglensing}, the effect of the lensing on the counts we fitted is smaller than 10\%. The model of lensing does not have free parameters. It is based on WMAP-7-years-best-fit cosmology and on some parameters taken at values given by the litterature. The lensing is thus not useful to reproduce the current observations, but is necessary to make predictions at bright fluxes ($>$100~mJy) in the sub-mm and mm range, where the effects of the lensing are large.\\

\section{Description of the model}

\subsection{Basic formulas}

The flux density $S_{\nu}$ at a frequency $\nu$ of a source lying at a redshift z is \citep{Hogg1999} is
\begin{equation}
S_{\nu} = \frac{(1+z) L_{(1+z)\nu}}{4 \pi D_L^2(z)} 
\end{equation}
where z is the redshift, $D_L$ is the luminosity distance of the source, and $L_{(1+z)\nu}$ is the luminosity at a frequency $(1+z) \nu$. The comoving volume corresponding to a redshift slice between z and z+dz and a unit solid angle is
\begin{equation}
\frac{dV}{dz} = D_H \frac{(1+z)^2 D_A^2}{\sqrt{\Omega_\Lambda + (1+z)^3 \Omega_m}}
\end{equation}
where $D_H$ is the Hubble distance ($D_H = c/H_0$), $D_A$ the angular distance to the redshift z. $\Omega_m$ and $\Omega_\Lambda$ are the normalized energy density of the matter and of the cosmological constant.

\subsection{Bolometric luminosity function and its evolution}

We assume that the luminosity function (LF) is a classical double exponential function \citep{Saunders1990}
\begin{equation}
\label{eq:lf}
\Phi(L_{IR}) = \Phi^\star \, \times \bigl ( \frac{L_{IR}}{L^\star})^{1-\alpha} \,  \times \,exp \bigl [ -\frac{1}{2\sigma^2} log_{10}^2(1+\frac{L_{IR}}{L^\star}))\bigl ]
\end{equation}
where $\Phi(L_{IR})$ is the number of sources per logarithm of luminosity and per comoving volume unit for an infrared bolometric luminosity $L_{IR}$. $\Phi_{\star}$ is the normalization constant characterizing the density of sources. $L_{\star}$ is the characteristic luminosity at the break. $1-\alpha$ and $1-\alpha-1/\sigma^2/ln^2(10)$ are the slope of the asymptotic power-law behavior at respectively low and high luminosity.\\

We assume a continuous evolution in luminosity and in density of the luminosity function with the redshift of the form $L^\star \propto (1+z)^{r_L}$  and $\Phi^\star \propto (1+z)^{r_\Phi}$ where $r_L$ and $r_\phi$ are coefficients driving the evolution in luminosity and density, respectively. It is impossible to reproduce the evolution of the LF with constant $r_L$ and $r_\phi$. We consequently authorize their value to change at some specific redshifts. The position of these breaks are the same for both $r_L$ and $r_\phi$. The position of the first redshift break is a free parameter and converge to the same final value for initial values between 0 and 2. To avoid a divergence at high redshift, we also add a second break fixed at z=2.

\begin{figure}
\centering
\includegraphics{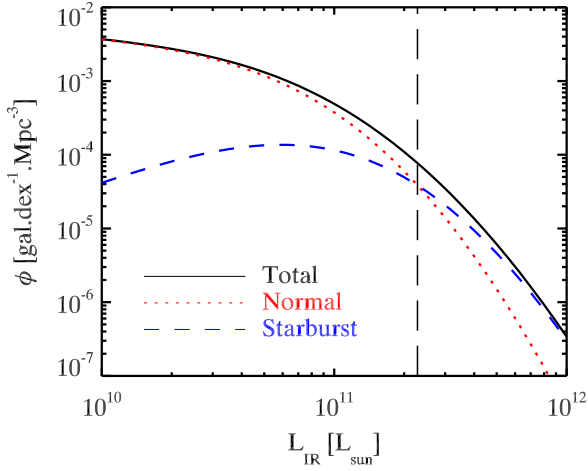}
\caption{\label{fig:lfbol_pops} \textit{Solid line}: Local infrared bolometric luminosity function from our best-fit model. \textit{Red dotted line}: contribution of the normal galaxies. \textit{Blue dashed line}: contribution of the starburst galaxies. \textit{Black vertical long dashed line}: luminosity of the transition between the two population ($L_{pop}$).}
\end{figure}

\subsection{Spectral energy distribution (SED) of the galaxies}

We use the \citet{Lagache2004} SED library. This library contains two populations: a starburst one and a normal one. This library is parametrized only by the infrared bolometric luminosity ($L_{IR}$). There is no evolution of the SED with the redshift. The normal population has a spectrum typical of spiral galaxy. The SED of this population does not evolve with $L_{IR}$. On the contrary, the starburst SED evolves with $L_{IR}$. The brighter the starburst galaxy, the hotter the dust.\\

The normal galaxies are dominant at low luminosity and the starburst at high luminosity. We thus chose arbitrary the following smooth function to describe the fraction of starburst galaxies as a function of the bolometric luminosity $L_{IR}$:
\begin{equation}
\label{eq:mixpop}
\frac{\Phi_{starburst}}{\Phi} = \frac{1+th \bigl [ log_{10}(L_{IR}/L_{pop}) /\sigma_{pop} \bigl ]}{2}
\end{equation}
where $th$ is the hyperbolic tangent function. $L_{pop}$ is the luminosity at which the number of normal and starburst galaxies are equal. $\sigma_{pop}$ characterizes the width of the transition between the two populations. At $L_{IR}=L_{pop}$, the fraction of starburst is 50\%. There are 88\% of starburst at $L_{IR}=L_{pop}\times10^{\sigma_{pop}}$, and 12\% at $L_{IR}=L_{pop}\times10^{-\sigma_{pop}}$.  The contribution of the different populations to the local infrared bolometric LF is shown in Fig. \ref{fig:lfbol_pops}.

\begin{table*}
\begin{tabular}{llr}
\hline
\hline
Parameter & Description & Value\\
\hline
$\alpha$ & Faint-end slope of the infrared bolometric LF & 1.223 $\pm$  0.044 \\
$\sigma$ & Parameter driving the bright-end slope of the LF & 0.406 $\pm$  0.019 \\
$L_\star$(z=0) ($\times 10^{10}~L_\odot$) & Local characteristic luminosity of the LF & 2.377 $\pm$  0.363 \\
$\phi_\star$ (z=0) ($\times 10^{-3}$ gal/dex/Mpc$^3$) & Local characteristic density of the LF & 3.234 $\pm$  0.266 \\
$r_{L_\star,lz}$ & Evolution of the characteristic  luminosity between 0 and $z_{break,1}$ & 2.931 $\pm$  0.119 \\
$r_{phi_\star,lz}$ & Evolution of the characteristic  density between 0 and $z_{break,1}$  &0.774 $\pm$  0.196 \\
$z_{break,1}$ & Redshift of the first break & 0.879 $\pm$  0.052 \\
$r_{L_\star,mz}$ & Evolution of the characteristic  luminosity between $z_{break,1}$ and $z_{break,2}$ &  4.737 $\pm$  0.301 \\
$r_{phi_\star,mz}$& Evolution of the characteristic  density of between $z_{break,1}$ and $z_{break,2}$ & -6.246 $\pm$  0.458 \\
$z_{break,2}$ &  Redshift of the second break  & 2.000 \, \,   (fixed) \\
$r_{L_\star,hz}$ &   Evolution of the characteristic luminosity for z$>z_{break,2}$ & 0.145 $\pm$  0.460 \\
$r_{phi_\star,hz}$ & Evolution of the characteristic density for z$>z_{break,2}$ &-0.919 $\pm$  0.651 \\
$L_{pop}$ ($\times 10^{10}~L_\odot$) & Luminosity of the transition between normal and starburst templates & 23.677 $\pm$  2.704 \\
$\sigma_{pop}$ & Width of the transition between normal and starburst templates & 0.572 $\pm$  0.056 \\
\hline
\end{tabular}
\caption{\label{tab:fitres} Parameters of our model fitted to our selection of infrared observations. The errors are derived from the MCMC analysis.}
\end{table*}

\subsection{Observables}

The number counts at different wavelengths are an essential constraint for our model. The source extraction biases are in general accurately corrected for these observables. The counts are computed with the following formula
\begin{eqnarray}
\frac{dN}{dS_\nu d\Omega} (S_\nu) = \\
\sum_{pop} \int_0^\infty  f_{pop}(L_{IR})\frac{dN}{dL_{IR} dV} \bigl |_{L_{IR}(S_\nu,z,pop)} \frac{dL_{IR}}{dS_\nu} \frac{dV}{dz d\Omega} dz \\
 = \sum_{pop} \int_0^\infty \frac{dN}{dS_\nu dz d \Omega} dz
\end{eqnarray}
where $dN/dS_\nu/d\Omega$ is the number of source per flux unit and per solid angle. $f_{pop}(L_{IR})$ is the fraction of the sources of a given galaxy population computed with the Eq. \ref{eq:mixpop}. $dN/dL_{IR}/dV$ is computed from the Eq. \ref{eq:lf}
\begin{equation}
\frac{dN}{dL_{IR} dV} = \frac{dN}{d log_{10}(L_{IR}) L_{IR} log(10) dV} = \frac{\Phi(L_{IR})}{L_{IR} log(10)}.
\end{equation}
$L_{IR}(S_\nu,z,pop)$ and $dL_{IR}/dS_\nu$ were computed on a grid in $S_\nu$ and z from the cosmology and the SED templates. These grids do not depend on the evolution of the LF nor on the population mixing parameters. These grids are thus generated only once and saved to accelerate the computation of the counts. Note that with this method, it is very easy to change the SED templates and/or add other populations.\\

Other measurements help to constraint our model. For example, the monochromatic luminosity $\Phi_{mono}$ function at a given redshift is
\begin{equation}
\Phi_{mono} = \sum_{pop} f_{pop} (L_{IR}(\nu L_\nu)) \phi(L_{IR}(\nu L_\nu)) \frac{d log_{10}(L_{IR})}{d (\nu L_\nu)}.
\end{equation}
We do not use the bolometric LFs, because they are biased by the choice of the assumed SED of the sources.\\

We can also compute the redshift distribution N(z) for a selection in flux $S_\nu > S_{\nu,cut}$ with
\begin{equation}
N(z,S_{cut}) = \int_{S_{\nu,cut}}^\infty \frac{dN}{dS_\nu dz} dS_\nu.
\end{equation}

The extragalactic background due to the galaxies at a given wavelength is
\begin{equation}
B_\nu = \int_{z=0}^\infty \int_{S_\nu = 0}^\infty S_\nu \frac{dN}{dS_\nu dz d\Omega} dS_\nu \, dz = \int_{S_\nu = 0}^\infty S_\nu \frac{dN}{dS_\nu d\Omega} dS_\nu
\end{equation}
and can be compared to the measurements of the CIB.\\

The level of the non-correlated fluctuations (shot-noise) of the CIB can be easily computed from our model with the equation:
\begin{equation}
P_{SN} = \int_0^{S_{\nu,cut}} S_\nu^2\frac{dN}{dS_\nu d\Omega} dS_\nu
\end{equation}
where $P_{SN}$ is the level of the non-correlated fluctuations and $S_{\nu,cut}$ the flux limit for the cleaning of the resolved sources.\\

\subsection{Effect of the strong lensing on the counts}

\begin{figure}
\centering
\includegraphics{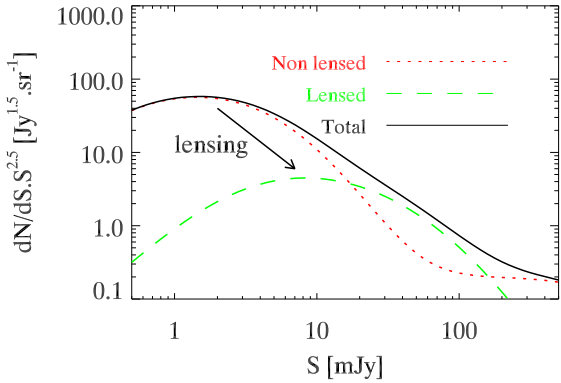}
\caption{\label{fig:lensing_effect} Effect of the lensing on the number counts at 850 microns. The contribution of lensed source is multiply by 10 to underline the effect of the lensing on the counts. \textit{Red dotted line}: counts of non-lensed sources. \textit{Green dashed line}: counts of lensed sources. \textit{Black solid line}: total counts.}
\end{figure}

We use a simple strong lensing model based on \citet{Perrotta2001,Perrotta2002}. It supposes that the dark matter halos are singular isothermal spheres. The cross-section $\sigma$ of a halo for a magnification $\mu$ larger than $\mu_{min}$ is
\begin{equation}
\sigma(\mu>\mu_{min}) = \frac{4 \pi \alpha^2 D_{A,ls}}{\mu^2}
\end{equation}
where $D_{A,ls}$ is the angular-diameter distance between the lens and the source and $\alpha$ is given by
\begin{equation}
\alpha = 4 \pi \frac{\sigma_v^2}{c^2}
\end{equation} 
where c is the speed of light and $\sigma_v$ the velocity dispersion in the halo, which depends on the cosmology, the redshift and the mass of the halo.\\

The probability $P(\mu_{min},z_s)$ for a source at a redshift $z_s$ to be magnified by a factor greater than $\mu_{min}$ is 
\begin{eqnarray}
P(\mu>\mu_{min},z_s) = \\
\frac{(1+z_s)^2}{4 \pi D_c(z_s)} \int_0^{z_s} \int_0^\infty \frac{dN}{d(log_{10}(M)) \, dV} \sigma(\mu > \mu_{min}) \frac{dV}{dz} dM dz
\end{eqnarray}
where $z_s$ is the redshift of the source, $D_c$ the comoving radial distance, $\frac{dN}{d(log_{10}(M)) \, dV}$ is the halo mass function, and $\frac{dV}{dz} $ is the comoving volume associated to the redshift slice dz. We use the halo mass function of \citet{Reed2007}.\\

The counts derived by our model take into account the fact that a small fraction of the sources are gravitationally magnified. The observed number counts taking into account the lensing $(dN/dS_\nu/d\Omega)_{lensed}$ are computed from initial counts $dS_\nu/dz/d\Omega$ with
\begin{eqnarray}
\bigl (\frac{dN}{dS_\nu d\Omega} \bigr)_{lensed}(S_\nu) = \\
\int_0^\infty \int_{\mu_{min}}^{\mu_{max}} \frac{dP}{d\mu}(z) \frac{1}{\mu}  \frac{dN}{dS_\nu dz d\Omega} \left( \frac{S_\nu}{\mu}, z \right ) d\mu \, dz.
\end{eqnarray}
Practically, this operation is performed multiplying the vector containing the counts for a given redshift slice by a matrix describing the effect of lensing. This lensing matrix has diagonal coefficients values around 1, and small ($<10^{-3}$) non-diagonal terms. These non-diagonal terms describe how magnified faint sources affect the counts at brighter flux. The effect on the monochromatic luminosity function was computed in the same way. We chose $\mu_{min}=2$ which corresponds to the limit of the validity of the strong-lensing hypothesis \citep{Perrotta2001}. The spatial extension of the lensed galaxies limits the maximum magnification. According to \citet{Perrotta2002}, $\mu_{max}$ is in the 10-30 range. We chose to use $\mu_{max}$=20 in this paper. \citet{Negrello2007} used $\mu_{min}$=2 and $\mu_{max}$=15.\\

Fig. \ref{fig:lensing_effect} illustrates how number counts are affected by lensing. This figure is based on the number counts predicted by the model at 850~$\mu$m with a probability of magnification multiplied by a factor 10 to better show this effect. The green dashed line is contribution of the lensed sources. Due to the magnification, the peak of this contribution is at higher flux than for non-lensed sources, and due to the small probability of lensing, the peak is lower than for non-lensed sources. This effect of the magnification on the counts become non negligible when the slope of the counts is very steep, like in the sub-mm and mm domain.

\section{Fitting the model parameters on the data}

Our model has several free parameters. We tried to have the minimum number of parameters. We determined them by fitting the model to published measurements of the counts and LFs. We used a Monte-Carlo Markov chain (MCMC) to find the best parameters, their uncertainties, and degeneracies. We do not fit the measured redshift distributions, because the cosmic variance and the selection effects are currently not enough accurately known.

\subsection{Data: extragalactic number counts}

\subsubsection{Data used for the fit}

We have chosen to fit the following data:
\begin{itemize}
\item \textit{Spitzer} MIPS counts of \citet{Bethermin2010a} at 24, 70 and 160~$\mu$m,
\item \textit{Herschel} SPIRE \citet{Oliver2010} counts at 250, 350 and 500~$\mu$m,
\item AzTEC counts of \citet{Austermann2010} and \citet{Scott2010} at 1.1~mm.
\end{itemize}

\subsubsection{Justification of our choice}

We fit only the differential number counts since the integral counts are highly correlated and the correlation matrix is rarely estimated.\\

The number counts were measured at numerous bands between 15~$\mu$m and 1.1~mm. We have chosen a collection of points. We were guided by the reliability of the measurements and their error bars.\\

Number counts at 15~$\mu$m based on the infrared space observatory (ISO) data \citep{Elbaz1999,Gruppioni2002} and on the Akari data \citep{Pearson2010,Hopwood2010}  exhibit a discrepancy by a factor of about 2, and their errors do not include cosmic variance. The results of these papers were not fitted. Nevertheless, we compared a posteriori to our results to check consistency in Sect. \ref{sect:comp_nofit}.\\

We fitted the \textit{Spitzer} MIPS counts of \citet{Bethermin2010a} at 24, 70 and 160~$\mu$m. These points were built from the data of FIDEL, COSMOS and SWIRE legacy programs. The errors bars take into account the cosmic variance. These counts agree with previous \textit{Spitzer} measurements of \citet{Papovich2004,Shupe2008,Le_Floch2009,Frayer2009} and \textit{Herschel} measurements of \citet{Berta2010} (in which the different fields are not combined).\\

 At 250~$\mu$m, 350~$\mu$m and 500~$\mu$m, we fitted \textit{Herschel} SPIRE \citet{Oliver2010} counts which take into account the cosmic variance and the deboosting uncertainty. These counts agree with the BLAST measurements of \citet{Patanchon2009} and \citet{Bethermin2010b} and the \textit{Herschel} measurements of \citet{Clements2010}. We chosen \citet{Oliver2010} counts because \textit{Herschel} data are better than BLAST ones and because \citet{Clements2010} counts use only Poissonian error bars, which could be largely underestimated. For instance, \citet{Bethermin2010a} estimate that the Poissonian uncertainties underestimate the real sample uncertainties by a factor 3 for counts around 100~mJy at 160~$\mu$m in a 10 deg$^2$ field.\\
 
We do not fit the 850~$\mu$m observation because of the large discrepancies between the submillimeter common-user bolometer array (SCUBA) observations \citep{Coppin2006} and the large APEX bolometer Camera (LABOCA) ones \citep{Weiss2009}. We discuss this problem in the Sect. \ref{sect:comp_nofit}.\\
 
We fitted the AzTEC measurements at 1.1~mm of \citet{Austermann2010} and \citet{Scott2010}. The area covered by AzTEC is small compared to \textit{Spitzer} and \textit{Herschel}. We used two independent measurements of the AzTEC counts to increase the weight of mm observations in our fit.\\
 
\subsection{Data: monochromatic luminosity functions}

\label{sect:data_lfs}

\subsubsection{Data used for the fit}

We have chosen to fit the following data:
\begin{itemize}
\item IRAS local luminosity function at 60~$\mu$m of \citet{Saunders1990}
\item \textit{Spitzer} local luminosity function at 24~$\mu$m of \citet{Rodighiero2009},
\item \textit{Spitzer} luminosity function at 15~$\mu$m at z=0.6 of \citet{Rodighiero2009},
\item \textit{Spitzer} luminosity function at 12~$\mu$m at z=1 of \citet{Rodighiero2009},
\item \textit{Spitzer} luminosity function at 8~$\mu$m at z=2 of \citet{Caputi2007},
\end{itemize}

\subsubsection{Justification of our choice}

We fitted some monochromatic luminosity functions. We chose only wavelengths and redshifts for which no K-corrections are needed. These observations strongly constrain the parameters driving the redshift evolutions of our model.\\

From the \citet{Rodighiero2009} LFs measured with the \textit{Spitzer} data at 24~$\mu$m, we computed 3 non K-corrected LFs at z=0, 0.6 and 1. We used their local LF at 24~$\mu$m. At z = 0.6 and 1, instead of using directly their results in their redshift bins, we combined their 15~$\mu$m LF at z=0.6 (respectively 12~$\mu$m LF at z=1) in the $0.45<z<0.6$ and $0.6<z<0.8$ bins (respectively $0.8<z<1.0$ and $1.0<z<1.4$) to obtain 15~$\mu$m LF at z=0.6 (respectively 12~$\mu$m LF at z=1). The error on a point is the maximum of the combination of the statistical errors of the two bins and of the difference between the measures in the two bins. The second value is often larger due to the quick evolution of the LF and the cosmic variance. We fitted only the points that do not suffer incompleteness to avoid possible biases. We also fitted the 8~$\mu$m at z=2 of \citet{Caputi2007}.\\

We also fitted the local LF at 60~$\mu$m determined from the infrared astronomical satellite (IRAS) data \citep{Saunders1990} to better constrain the faint-end slope of the local LF. Due to the strong AGN contamination at 60~$\mu$m in the ULIRG regime, we did not fit the points brighter than 10$^{11.5} \, L_\odot$ at 60~$\mu$m.\\

\subsection{Data: CIB}

The bulk of the CIB is not resolved at SPIRE wavelengths. We thus used the absolute measurement of the CIB level in SPIRE bands as a constraint for our model. We used the \citet{Lagache1999} measurement performed on the far-infrared absolute spectrophotometer (FIRAS) data: 11.7$\pm$2.9 nW.m$^2$.sr$^{-1}$ at 250~$\mu$m, 6.4$\pm$1.6 nW.m$^2$.sr$^{-1}$ at 350~$\mu$m and 2.7$\pm$0.7 nW.m$^2$.sr$^{-1}$ at 500~$\mu$m. We assume that the CIB is only due to galaxies and thus neglect a possible extragalactic diffuse emission. 

\subsection{Calibration uncertainties}

\begin{table}
\centering
\begin{tabular}{lrr}
\hline
\hline
Instrument & Calibration parameter ($\gamma_b$) & Calib. uncertainty\\
\hline
MIPS 24~$\mu$m & 1.00$\pm$0.03 & 4\% \\
MIPS 70~$\mu$m & 1.06$\pm$0.04 & 7\% \\
MIPS 160~$\mu$m & 0.96$\pm$0.03 & 12\% \\
SPIRE 250~$\mu$m & 0.88$\pm$0.05 & 15\%\\
SPIRE 350~$\mu$m & 0.97$\pm$0.07 & 15\%\\
SPIRE 500~$\mu$m & 1.17$\pm$0.1 & 15\%\\
AzTEC 1.1~mm & 0.98$\pm$0.09 & 9\% \\
\hline
\end{tabular}
\caption{\label{table:calib} Calibration parameters and 1-$\sigma$ marginalized errors from our MCMC fit compared with calibration uncertainties given by the instrumental teams.}
\end{table}

The calibration uncertainty is responsible for correlated uncertainties between points measured at a given wavelength with the same instrument. A change in the calibration modifies globally the number counts and the LF. Assuming the "good" calibration is obtained in multiplying the fluxes by a factor $\gamma$, the "good" normalized counts are obtained with $S_{new} = \gamma S$ and $(S_{good}^{2.5}dN/dS_{good}) = \gamma^{1.5} (S^{2.5}dN/dS)$. The effect on the LF in dex per volume unit is more simple. We just have to shift the luminosity in abscissa by a factor $\gamma$.\\

We added to our free parameters a calibration parameter for each fitted band (see Table \ref{table:calib}). We took into account the uncertainties on the calibration estimated by the instrumental team in our fit \citep{Stansberry2007,Gordon2007,Engelbracht2007,Swinyard2010,Scott2010}.

\subsection{Fitting method}

To fit our points, we assumed that the uncertainties on the measurements and on the calibrations are Gaussian and not correlated. The log-likelihood is then
\begin{equation}
-log(L(\theta)) = \sum_{k=1}^{Npoints} \frac{(m_k-m_{model,k}(\theta))^2}{2 \sigma_m^2} + \sum_{b=1}^{Nband} \frac{(\gamma_b-1)^2}{2 \sigma_{calib,b}^2}
\end{equation}
where L is the likelihood, $\theta$ the parameters of the model, $m_k$ a measurement, $m_{model,k}$ the prediction of the model for the same measurement, $\sigma_m$ the measurement uncertainty on it, $\gamma_b$ the calibration parameter of the band b and $\sigma_{calib,b}$ the calibration uncertainty for this band.\\

We used a Monte Carlo Markov chain (MCMC) Metropolis-Hastings algorithm \citep{Chib1995,Dunkley2005} to fit our model. The method consists in a random walk in the parameter space. At each step, a random shift of the parameters is done using a given fixed proposal density. A step n is kept with a probability of 1 if $L(\theta_n)>L(\theta_{n-1})$ or with a probability $L(\theta_n)/L(\theta_{n-1})$ else. The distribution of the realization of the chain is asymptotically the same as the underlying probability density. This property is thus very convenient to determine the confidence area for the parameters of the model.\\

We used the Fisher matrix formalism to determine the proposal density of the chain from initial parameters values set manually. The associated Fisher matrix is:
\begin{equation}
F_{ij}(\theta) = \sum_{k=1}^{Npoints} \frac{\partial m_{model,k}}{\partial \theta_i}  \frac{\partial m_{model,k}}{\partial \theta_j}  \frac{1}{2 \sigma_m^2} \left (+ \frac{1}{2 \sigma_{calib,b}^2} \right )
\end{equation}
where $\theta$ is a vector containing the model and calibration ($\gamma_b$) parameters. The term in brackets appears only for the diagonal terms corresponding to a calibration parameter. We ran a first short chain (10~000 steps) and computed a new proposal density with the covariance matrix of the results. We then ran a second long chain of 300~000 steps. The final chain satisfies the \citet{Dunkley2005} criteria ($j^\star>20$ and $r<0.01$).

\section{Results of the fit}

\begin{figure*}
\centering
\includegraphics[width=17.5cm]{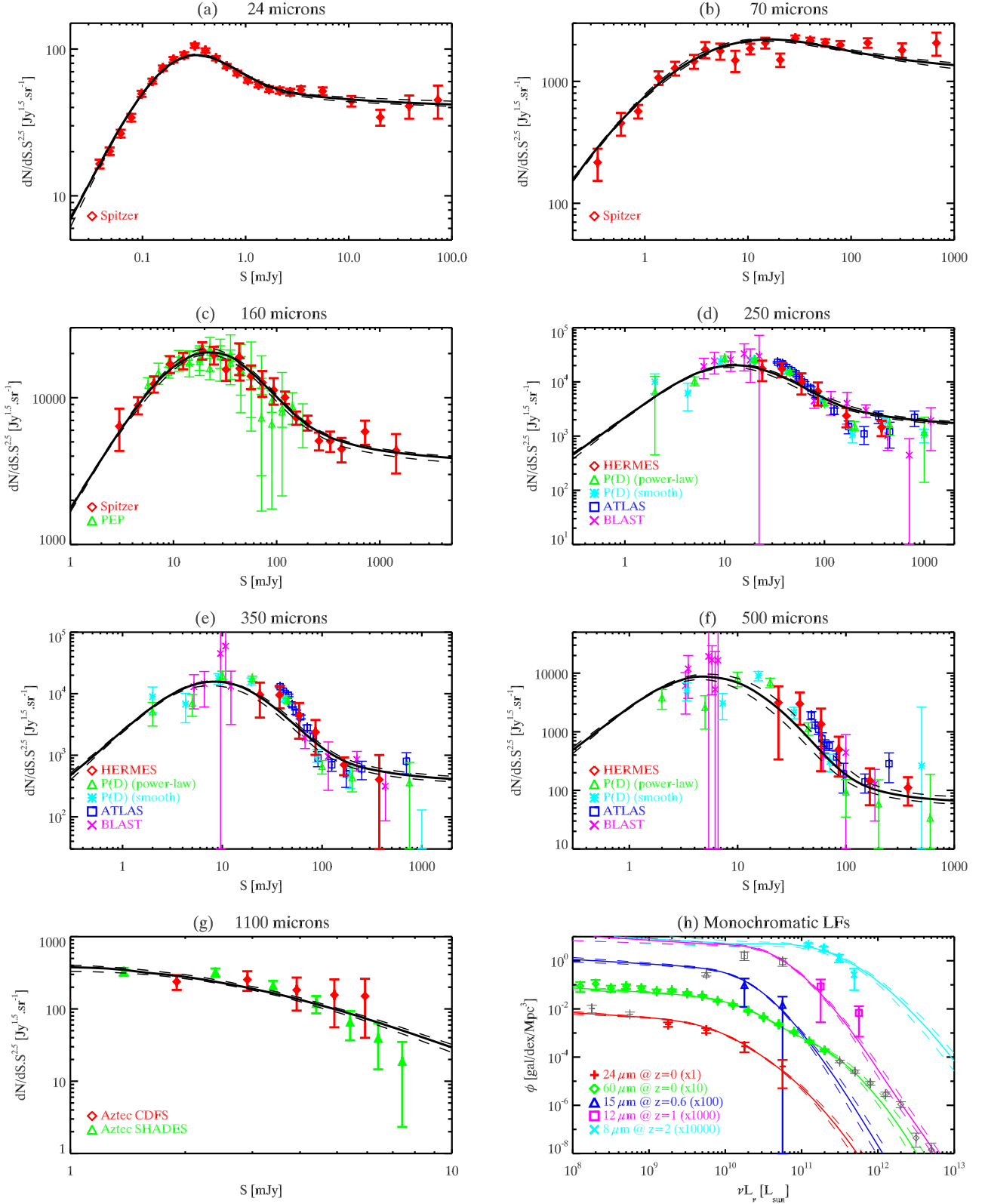}

\caption{\label{fig:fitres} (a) to (f): Differential extragalactic number counts used for the fit. (h): Monochromatic LFs at different wavelengths and redshifts. (a) to (h): The fitted points are thicker. \textit{Black solid line}: our best-fit model. \textit{Black dashed line}: 1-$\sigma$ range of the model. (a) to (c): \textit{Red diamonds}: \citet{Bethermin2010a} \textit{Spitzer} legacy number counts. (c): \textit{Green triangles}: \citet{Berta2010} \textit{Herschel}/PEP number counts. (d) to (f): \textit{Red diamonds}: \citet{Oliver2010} \textit{Herschel}/Hermes number counts. \textit{Green triangles}: \citet{Glenn2010} \textit{Herschel}/Hermes P(D) analysis. \citet{Clements2010} \textit{Herschel}/ATLAS number counts. \textit{Purple cross}: \citet{Bethermin2010b} BLAST number counts. (g): \textit{Green triangles}: \citet{Scott2010} AzTEC number counts in the CDFS field. \textit{Green triangles}: \citet{Austermann2010} AzTEC number counts in the SHADES field. (h): \textit{Red plus}: \citet{Rodighiero2009} local 24~$\mu$m LF (not fitted points in grey). \textit{Green diamonds}: \citet{Saunders1990} local 60~$\mu$m LF (shifted by a factor 10 on the y-axis; not fitted points in grey); \textit{Blue triangles}: \citet{Rodighiero2009} 15~$\mu$m LF at z=0.6 (shifted by a factor 100 on the y-axis; not fitted points in grey). \textit{Purple squares}: \citet{Rodighiero2009} 12~$\mu$m LF at z=1 (shifted by a factor 1000 on the y-axis; not fitted points in grey). \textit{Cyan crosses}: \citet{Caputi2007} 8~$\mu$m LF at z=2 (shifted by a factor 10~000 on the y-axis).}

\end{figure*}

\subsection{Quality of the fit}

Our final best-fit model have a $\chi^2$ ($\chi^2 = -2 log(L)$ because all errors are assumed to be Gaussian) of 177 for 113 degrees of freedom. Our fit is thus reasonably good. The parameters found with the fit are given in Table \ref{tab:fitres} (the uncertainties are computed from the MCMC). The calibration factor are compatible with the calibration uncertainties given by the instrumental teams with a $\chi^2$ of 2.89 for 7 points (see Table \ref{table:calib}). The results are plotted in Fig. \ref{fig:fitres}.

\subsection{Comparison between the model and the observed counts used in the fit}

The \citet{Bethermin2010a} points fit globally well, with some exceptions. Our model is lower by about 15\% than two points around 300~$\mu$Jy at 24~$\mu$m. These two points are built combining the FIDEL, COSMOS and SWIRE fields. The SWIRE fields are shallow fields and the counts could be affected by the Eddington bias. We also observe a slight under-prediction of the bright ($S_{70}>$50~mJy) counts at 70~$\mu$m. We also plotted the \citet{Berta2010} counts at 160~$\mu$m measured using the photodetector array camera and spectrometer (PACS) on the \textit{Herschel} satellite. These counts agree with \citet{Bethermin2010a} and our model.\\

Our model fits globally well the \citet{Oliver2010} and \citet{Bethermin2010b} counts, excepting a slight under-prediction of the counts between 30~mJy and 100~mJy at 500~$\mu$m. There is a mild disagreement with the \citet{Clements2010} counts, but their errors bars do not take into account the cosmic variance and are thus underestimated. We also plotted the results of the P(D) analysis of \citet{Glenn2010}. These points and especially the error bars must be interpreted with caution (see the complete discussion in \citet{Glenn2010}). We have plotted the knots of the smooth and power-law models. They globally agree with our model.\\

Our model agrees very well with the AzTEC counts of \citet{Austermann2010} and \citet{Scott2010}. The contribution of the strong lensing objects to the AzTEC counts is weak ($<$10\%, see Sect. \ref{section:stronglensing}).

\subsection{Comparison between the model and the observed monochromatic LFs}

Our model fits well our collection of LFs \citep{Saunders1990,Caputi2007,Rodighiero2009}, excepting the brightest point of \citet{Caputi2007}. In Fig. \ref{fig:fitres}, we arbitrary shifted the different LFs on the y-axis to obtain a clearer plot. Our model underestimates the 60~$\mu$m local LF in the ULIRG regime. It is expected because our model do not contain AGNs and confirms our choice of not fitting these points (Sect. \ref{sect:data_lfs}).  

\subsection{Comparison between the model and the observed counts not used in the fit}

\label{sect:comp_nofit}

\begin{figure}
\centering
\includegraphics{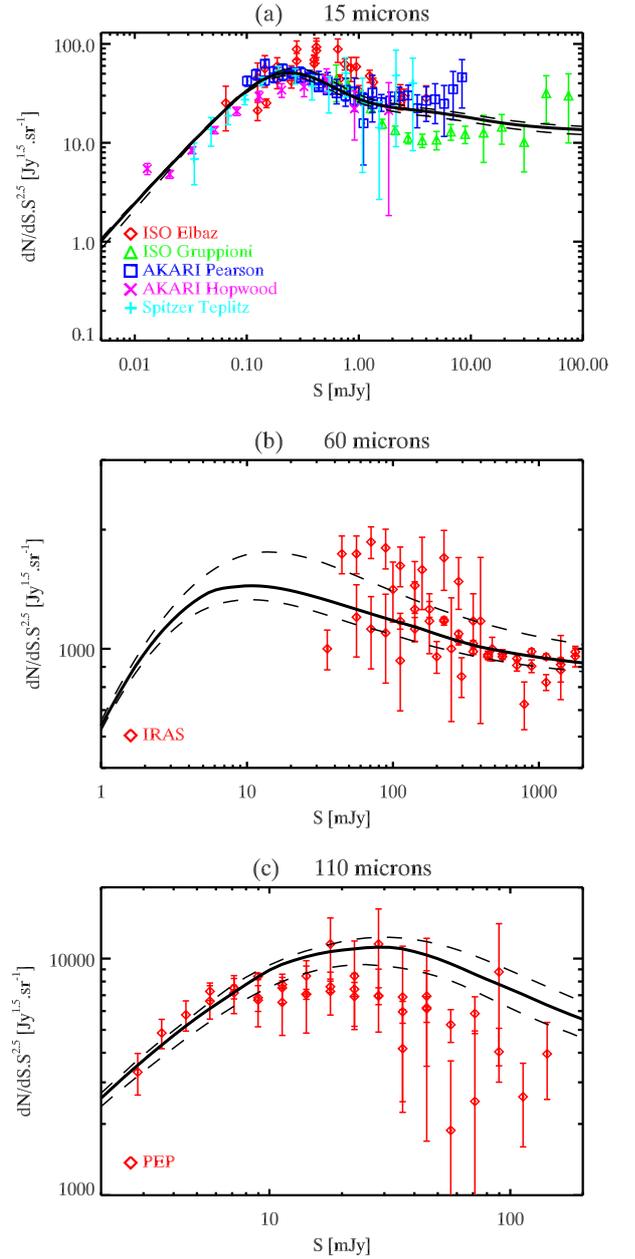}

\caption{\label{fig:othercounts} (a) to (c): Differential extragalactic number counts not used for the fit. \textit{Black solid line}: our best-fit model. \textit{Black dashed line}: 1-$\sigma$ range of the model. (a): \textit{Red diamonds}: \citet{Elbaz1999} ISO counts. \textit{Green triangle}: \citet{Gruppioni2002} ISO counts. \textit{Blue squares}: \citet{Pearson2010} Akari counts. \textit{Purple cross}: \citet{Hopwood2010} Akari (lensed) counts. \textit{Cyan plus}: \citet{Teplitz2010} \textit{Spitzer}/IRS counts. (b): \textit{Red diamonds}: \citet{Hacking1987}, \citet{Lonsdale1990}, \citet{Rowan1990}, \citet{Saunders1990}, \citet{Gregorich1995} and \citet{Bertin1997} IRAS counts. (c): \textit{Red diamonds}: \citet{Berta2010} \textit{Herschel}/PEP counts.}
\end{figure}

\begin{figure}
\centering
\includegraphics{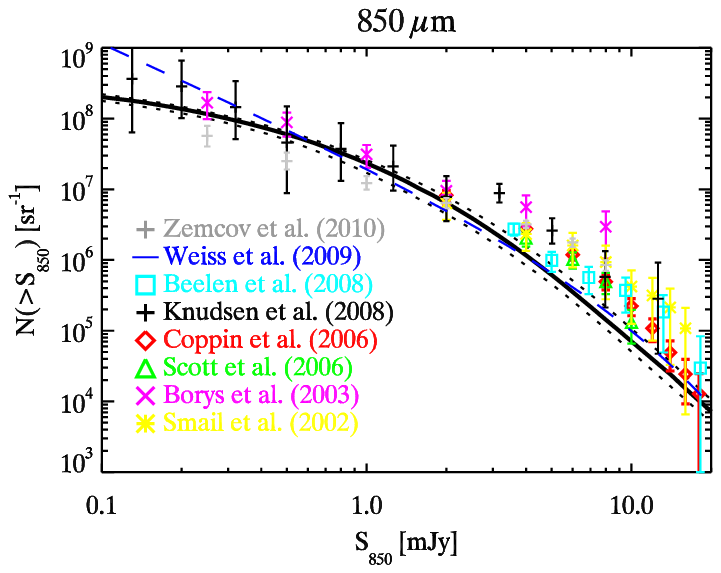}
\caption{\label{fig:ic850} Integral number counts at 850~$\mu$m.  \textit{Black solid line}: our best-fit model. \textit{Black dashed line}: 1-$\sigma$ range of the model. \textit{Grey plus}: \citet{Zemcov2010} combined SCUBA lensed counts. \textit{Blue dashed line}: \citet{Weiss2009} LABOCA P(D) (Schechter model). \textit{Red diamonds}: \citet{Coppin2006} SCUBA SHADES counts. \textit{Cyan square}: \citep{Beelen2008} LABOCA counts around the J2142-4423 Ly$\alpha$ protocluster. \textit{Black plus}: \citet{Knudsen2008} combined SCUBA lensed counts; \textit{Green triangles}: \citet{Scott2006} revisited SCUBA counts. \textit{Purple cross}: \citet{Borys2003} SCUBA HDFN counts. \textit{Yellow asterisks}: \citet{Smail2002} lensed counts.}
\end{figure}

We also compared our results with the counts at other wavelengths. They are plotted on Fig. \ref{fig:othercounts} and \ref{fig:ic850}. The 1-$\sigma$ region of the model includes the $\gamma_b$ uncertainty of Akari at 15~$\mu$m (4\%, \citet{Ishihara2010}), PACS at 110~$\mu$m (about 10\%, \citet{Berta2010}) and LABOCA at 850~$\mu$m (8.5\%, \citet{Weiss2009}). The uncertainty on $\gamma_b$ is about the same for LABOCA and SCUBA ($\sim$10\%, \citet{Scott2006}). The uncertainties on the model are larger at these non-fitted wavelengths because the correlations between the model and the calibration parameters are not taken into account by the fit.\\

At 15~$\mu$m, the \citet{Elbaz1999} counts from different fields are not compatible between them, but our counts pass in the cloud of points. The \citet{Gruppioni2002} counts are significantly lower than our model and other works. We marginally agree with the \citet{Pearson2010} counts. The \citet{Hopwood2010} counts measured with Akari in a field around Abell 2218 are lower than our model by about 25\%. Nevertheless, their field is very narrow and their estimation may suffer from cosmic variance. Finally, we well agree with the very recent \citet{Teplitz2010} measurements performed with the infrared spectrograph (IRS) onboard the \textit{Spitzer} space telescope.\\

We compare our counts to \citet{Hacking1987}, \citet{Lonsdale1990}, \citet{Rowan1990}, \citet{Saunders1990}, \citet{Gregorich1995} and \citet{Bertin1997} at 60~$\mu$m from IRAS data. There are disagreements between the different observations and some error bars may be underestimated, but our model globally agrees with the cloud of points.\\

We can also compare the prediction of our model with \citet{Berta2010} counts at 110~$\mu$m. Our model globally agrees with their work. Nevertheless, our model tends to be higher than their measurement near 100~mJy. Observations on several larger fields will help to see if this effect is an artifact or not.\\

At 850~$\mu$m, we very well agree with the P(D) analysis of the LABOCA data of \citet{Weiss2009} (see Fig. \ref{fig:ic850}). But, the measurements performed with SCUBA \citep{Borys2003,Scott2006,Coppin2006} and LABOCA \citep{Beelen2008} are significantly higher than our model at 6 and 8~mJy. At low flux ($<$2~mJy), our model agrees very well with the measurement performed in lensed region \citep{Smail2002,Knudsen2008,Zemcov2010}.\\

\begin{figure}
\centering
\includegraphics{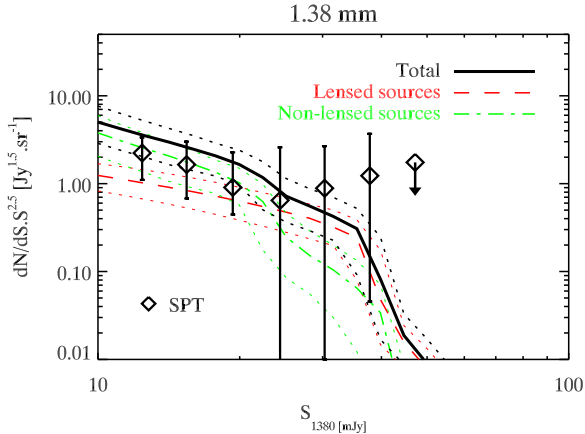}
\caption{\label{fig:spt_counts} Number counts at 1.38~mm of dusty sources without IRAS 60~$\mu$m counterpart . \textit{Black diamonds}: \citet{Vieira2009} south pole telescope (SPT) measurements. \textit{Black solid line}: Total contribution of $S_{60}<0.2$~Jy sources. \textit{Green dot-dashed line}: Contribution of the non-lensed sources. \textit{Red dashed line}: Contribution of the strongly-lensed sources. \textit{Dotted lines} 1-$\sigma$ contours.}
\end{figure}

We also compare our model predictions with SPT measurements at 1.38~mm \citep{Vieira2009}. At this wavelength, the contribution of the synchrotron emission of the local radio-galaxies to the counts is not negligible. Nevertheless, these sources can be separated from dusty galaxies considering their spectrum. We thus compare our results with the counts of dusty sources. \citet{Vieira2009} measured counts for all the dusty sources and for the dusty sources without IRAS 60~$\mu$m counterpart. Our model agrees with these two measurements. Fig. \ref{fig:spt_counts} shows the counts of the non-IRAS dusty sources. The 7.2\% calibration uncertainty of SPT is taken into account in the 1-$\sigma$ region of the model.

\subsection{Comparison with the observed redshift distributions}

\begin{figure}
\centering
\includegraphics{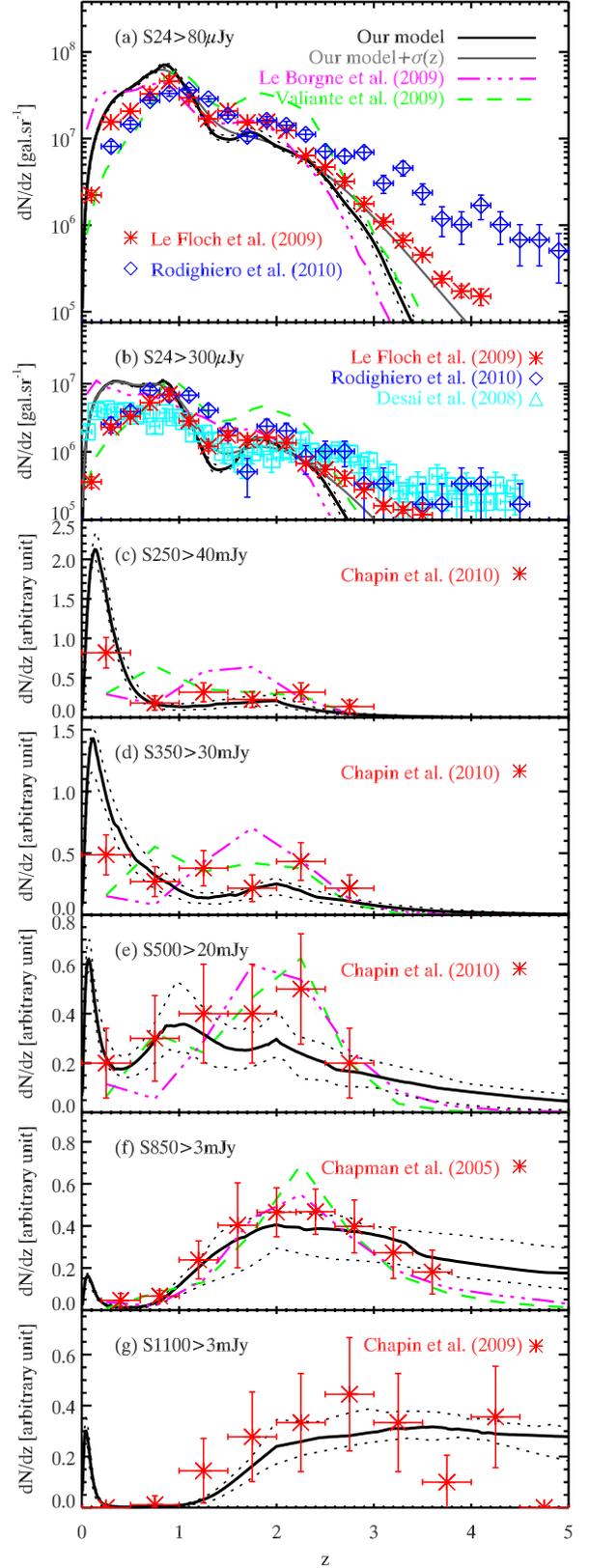}
\caption{\label{fig:nz} Redshift distribution of the S$_{24}>$80~$\mu$Jy (a),  S$_{24}>$300~$\mu$Jy (b), S$_{250}>$40~mJy (c), S$_{350}>$30~mJy (d), S$_{500}>$20~mJy (e), S$_{850}>$3~mJy (f), S$_{1100}>$3~mJy (g) sources. These measurements are not fitted. \textit{Black solid line}: our best-fit model.  \textit{Black dotted line}: 1-$\sigma$ range of the model. \textit{Grey solid line}: our best-fit model convolved by a gaussian of $\sigma_z=0.125 z$. \textit{Purple three dot-dashed line}: \citet{Le_Borgne2009} model. \textit{Green dashed line}: \citet{Valiante2009} model. \textit{Red asterisks}: \citet{Le_Floch2009} (a, b), \citet{Chapin2010} (c, d, e), \citet{Chapman2005} (f) and \citet{Chapin2009} (g) measurements. \textit{Blue diamonds}: \citet{Rodighiero2009} measurements(a, b). \textit{Cyan squares}: \citet{Desai2008} measurements (b).}
\end{figure}

In Fig. \ref{fig:nz}, we compare our model predictions with observed redshift distributions. At 24~$\mu$m, our model over-predicts by about 20\% the number of sources below z=1 compared to \citet{Le_Floch2009} observations for the selection S$_{24}>80 \mu$Jy. Nevertheless, they exclude $i_{AB}^+$$<$20 galaxies and their number of sources at low redshift is thus underestimated. Our model also underpredicts the number of sources at z$>$3. But, the redshifts of the z$>$2 sources are only moderately accurate ($\sigma_z \approx 0.25$ for $i_{AB}^+>$25 at z$\sim$2). Because of the strong slope of the redshift distribution, a significant number of sources measured near z=3.5 could be sources lying around z=3 with overestimated redshift. If we convolve our model with a gaussian error with $\sigma_z=0.125 z$ to simulate the redshift uncertainties, the model and the measurements agrees (Fig. \ref{fig:nz}). The \citet{Le_Borgne2009} model agrees very well with the measurements, excepting at z$<$0.5 and z$>$2.5. The \citet{Valiante2009} model poorly reproduces this observation. The same observables was measured by \citet{Rodighiero2009}. Their results are in agreement with \citet{Le_Floch2009}, excepting at z$>$3, where they are higher. It could be explain by a larger $\sigma_z$ at high redshift.\\

We also compare the model with the redshift distribution of $S_{24}>300 \mu$Jy sources measured by \citet{Le_Floch2009}, \citet{Rodighiero2009} and \citet{Desai2008}. These different measurements exhibit disagreements below z=0.5. This difference could be explained by the removing of the brightest optical sources (see previous paragraph). Our model overestimates the number of sources at z$<$0.5 by a factor 2. There is a rather good agreements between the models and the measurements between z=0.5 and z=2.5, except a small overestimation by \citet{Valiante2009} near z=2. At higher redshifts, the measurements are significantly higher than the models. It could be explained by two reasons: an effect of the redshift uncertainties and the absence of AGN in our model.\\

We compare with the \citet{Chapin2010} redshift distributions of the BLAST isolated sources at 250~$\mu$m, 350~$\mu$m and 500~$\mu$m. This selection of isolated sources does not allow to know the effective size of the field. We thus normalized our model and the measured counts to have $\int dN/dz dz = 1$. Our predicted redshift distribution globally fits the measurements, except at low z at 250~$\mu$m and 350~$\mu$m. This difference could be explained by the selection of isolated sources, which could miss sources in structures at low redshift. The other models \citep{Le_Borgne2009,Valiante2009} underpredicts the number of sources at low z. \citet{Valiante2009} also slightly overpredicts the number of sources at z$\sim$1.5.\\

We compared the redshift distribution of the SCUBA sources at 850~$\mu$m with the prediction of our model. We use the selection-corrected measurements of \citet{Chapman2005} used by \citet{Marsden2010}. All the models agrees with this measurement.\\

We also compared the prediction of our model with the redshift distribution of the sources detected at 1.1~mm by AzTEC \citep{Chapin2009}. A significant part of the sources detected at this wavelength (10 over 28) are not identified, and the selection is not performed in flux, but in signal-to-noise ratio. Consequently, the normalization of the redshift distribution is not known. We thus use the same normalization than for the BLAST redshift distributions ($\int dN/dz dz = 1$). The behavior predicted by our model agrees well with the observations.\\

\begin{figure}
\centering
\includegraphics{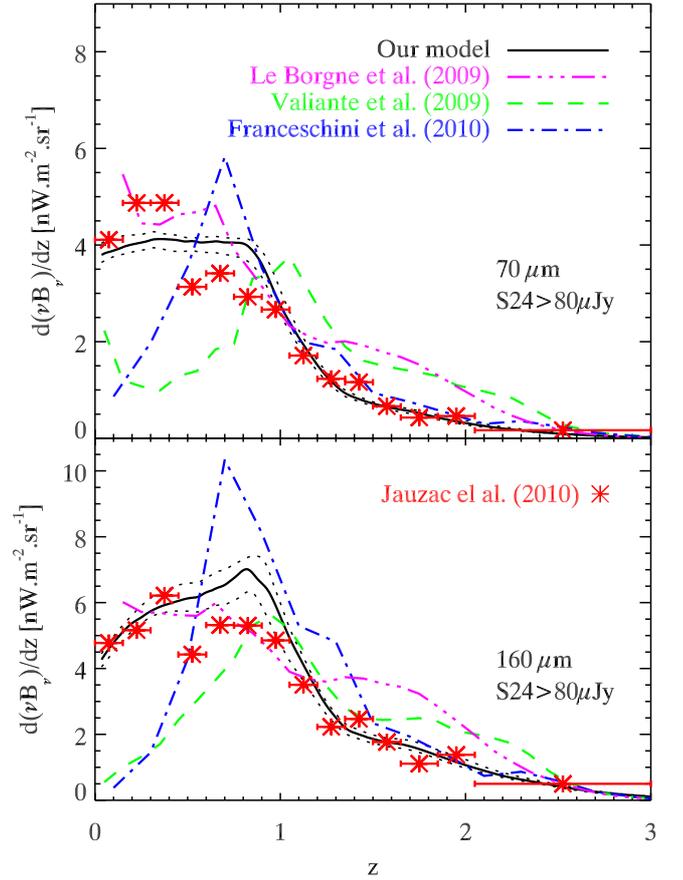}
\caption{\label{fig:dbdz} Differential contribution of the S$_{24}>80$~$\mu$Jy sources to the CIB as a function of the redshift at 70~$\mu$m (upper panel) and 160~$\mu$m (lower panel). \textit{Red asterisks}: measurement by stacking in the COSMOS field \citep{Jauzac2010}. \textit{Black solid line}: Our model (1-$\sigma$ limit in \textit{black dotted line}). \textit{Purple three dot-dashed line}: \citet{Le_Borgne2009} model. \textit{Green dashed line}: \citet{Valiante2009} model. \textit{Blue dot-dashed line}: \citet{Franceschini2009} model.}
\end{figure}

Recently, \citet{Jauzac2010} has measured the contribution of the S$_{24}>80$~$\mu$Jy to the CIB at 70 and 160~$\mu$m as a function of the redshift. Their stacking analysis allows to check the total far-infrared (FIR) emissions of the faint sources not resolved at these wavelengths. Our model agrees well with their results, except near z=0.5 (see Fig. \ref{fig:dbdz}), where their low data points could come from a large-scale underdensity in the COSMOS field at this redshift. The \citet{Le_Borgne2009} model overpredicts the contribution of the 24~$\mu$m sources at z$>$1. The \citet{Valiante2009} model does not reproduce the trend of these measurements. \citet{Franceschini2009} underestimate the contribution of the local sources and overestimate the contribution of z$\sim$1 sources.

\subsection{Comparison with the measured Poisson fluctuations of the CIB}

\begin{table*}
\centering
\begin{tabular}{llrrrr}
\hline
\hline
wavelength & reference & S$_{cut}$ & $P_{SN,mes}$  & $P_{SN,model}$ & $<$z$_{model}>$\\
$\mu$m & & mJy & Jy$^2$.sr$^{-1}$ & Jy$^2$.sr$^{-1}$ & \\
\hline
60 & \citet{Miville2002} & 1000 & 1600$\pm$300 & 2089$\pm$386 & 0.20$\pm$0.01 \\
90 & \citet{Matsuura2010} & 20 & 360$\pm$20 & 848$\pm$71 & 0.79$\pm$0.03 \\
100 & \citet{Miville2002} & 700 & 5800$\pm$1000 & 7364$\pm$1232 & 0.38$\pm$0.03 \\
160 & \citet{Lagache2007} & 200 & 9848$\pm$120 & 10834$\pm$3124 & 0.73$\pm$0.06 \\
250 & \citet{Viero2009} & 500 & 11700$\pm$400 & 11585$\pm$2079 & 0.81$\pm$0.08 \\
350 & \citet{Viero2009} & 400 & 6960$\pm$200 & 5048$\pm$1083 & 1.17$\pm$0.12 \\
500 & \citet{Viero2009} & - & 2630$\pm$100 & 1677$\pm$484 & 1.59$\pm$0.21 \\
%250 & \citet{Amblard2010} & 50 & 6100$\pm$120 & 6726$\pm$1207 & 1.09$\pm$0.06 \\
%350 & \citet{Amblard2010} & 50 & 4600$\pm$70 & 4446$\pm$1320 & 1.32$\pm$0.11 \\
%500 & \citet{Amblard2010} & 50 & 1800$\pm$80 & 1369$\pm$537 & 1.85$\pm$0.21 \\
1363 & \citet{Hall2010} & 15 & 17$\pm$2 & 10$\pm$3 & 4.07$\pm$0.24 \\
\hline
\end{tabular}
\caption{\label{tab:fluctuations} Level of the non-correlated fluctuations of the CIB at different wavelengths and comparison with the predictions of the model. The uncertainties on the model predictions take into account the uncertainties on $\gamma_b$. The mean redshift $<$z$_{model}>$ of the contribution to the fluctuations is a prediction of the model.}
\end{table*}

Table \ref{tab:fluctuations} summarizes the recent measurements of the non-correlated fluctuations of the CIB (P$_{SN}$) and the predictions of our model. Note that P$_{SN}$ depends strongly on the $S_{\nu,cut}$, the flux density at which the resolved sources are cleaned. We agree with the measurements of \citet{Miville2002} at 60~$\mu$m and 100~$\mu$m, \citet{Lagache2007} at 160~$\mu$m and \citet{Viero2009} at 250~$\mu$m and 350~$\mu$m. We found a value 35\% lower than \citet{Viero2009} at 500~$\mu$m. This is consistent with the slight under-estimation of the counts at 500~$\mu$m by our model. Our model is also about 40\% lower than the SPT measurements at 1.36~mm \citep{Hall2010}. It could be due to a lack of faint sources at high redshift in our model. We also disagree with \citet{Matsuura2010} at 90~$\mu$m within a factor of 2. Nevertheless, they cleaned all the detected sources without fixed cut in flux. We took their "mean" value of 20~mJy for the flux cut. The high sensitivity of the measurements of the flux cut could thus explains this difference (for instance, a decrease of the flux cut by 25\% implies a decrease of the fluctuations of 19\%).\\

We also computed the mean redshift at which the fluctuations are emitted with
\begin{equation}
<z> = \frac{\int_{0}^{\infty} z \frac{dP_{SN}}{dz} dz}{\int_{0}^{\infty} \frac{dP_{SN}}{dz} dz}.
\end{equation}
The results are written in Table \ref{tab:fluctuations}. As expected, the mean redshift increases with the wavelength. Studying the long wavelengths is thus very useful to probe high redshift populations.

\subsection{Comparison with the pixel histogram of the BLAST maps}

\label{sect:pd}

The quality of our counts at low fluxes in the sub-mm range can be tested using a P(D) analysis \citep{Condon1974,Patanchon2009,Glenn2010}. Without instrumental noise, the probability density of the signal in a pixel of the map, P(D), is given by:
\begin{equation}
P(D) = \int_0^\infty \left [  exp \left ( \int_0^\infty R(x) e^{i \omega x} dx - \int_0^\infty R(x) dx  \right ) \right ] e^{-i \omega D} d\omega 
\end{equation}
where R(x) is defined by
\begin{equation}
R(x) = \int \frac{1}{b} \frac{dN}{dS_\nu}\left(\frac{x}{b}\right) d\Omega.
\end{equation}
This probability distribution must be convolved by the distribution of the instrumental noise. We also subtract the mean of this distribution.\\

\begin{figure}
\centering
\includegraphics{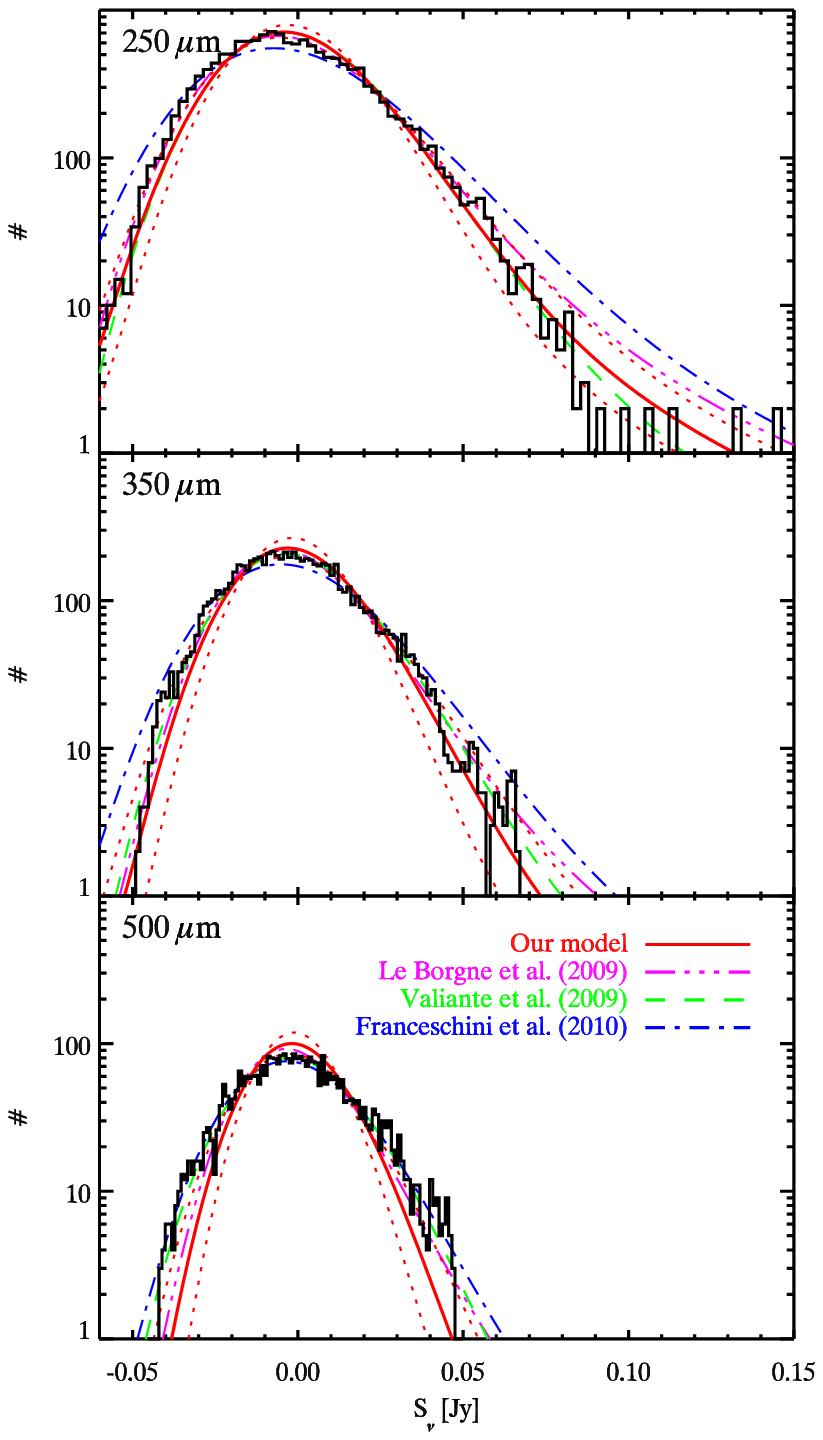}
\caption{\label{fig:pd_blast} Comparison with the BLAST pixel histogram at 250~$\mu$m (\textit{upper panel}), 350~$\mu$m (\textit{middle panel}) and 500~$\mu$m (\textit{lower panel}). (\textit{Black histogram}): histogram of the values of the central part of the BLAST beam-smoothed map in Jy/beam. (\textit{Red solid line}): distribution predicted by our model using a P(D) analysis. Our analysis does not include the clustering. \textit{Purple three dot-dashed line}: \citet{Le_Borgne2009} model. \textit{Green dashed line}: \citet{Valiante2009} model. \textit{Blue dot-dashed line}: \citet{Franceschini2009} model.}
\end{figure}

We tested our model with the deepest part of the observations of the CDFS by the BLAST team. We kept only the pixels of the map with a coverage larger than 90\% of the maximum coverage. We smoothed the signal, noise and beam map by a gaussian kernel with the same full width at half maximum than the BLAST beam. This smoothing reduces the effect of the instrumental noise \citet{Patanchon2009}. The predictions of our model and the BLAST pixel histograms at 250~$\mu$m, 350~$\mu$m and 500~$\mu$m are shown in Fig. \ref{fig:pd_blast}. The uncertainties on the model predictions take into account the BLAST calibration uncertainties \citep{Truch2009}. The model agrees rather well with the data. Nevertheless the measured histogram is slightly larger than the predictions of the model, especially at 500~$\mu$m. It is consistent with the slight under-estimation by our model of the counts at 500~$\mu$m (the higher the counts, the larger the histogram). The clustering of the galaxies (negliged in this analysis) tends to enlarge the histogram of about 10\% and could also contributes to this disagreement \citep{Takeuchi2004,Patanchon2009,Glenn2010}. The \citet{Valiante2009} model fits very well fit the BLAST pixel histograms. \citet{Le_Borgne2009} and \citet{Franceschini2009} overpredicts the number of bright pixels at 250~$\mu$m and 350~$\mu$m (S$_\nu>$50mJy). It is consistent with the fact that they overpredict the counts at high flux \citep{Oliver2010,Glenn2010}.

\subsection{Degeneracies between parameters}

The Pearson correlation matrix of our model is given in Tab. \ref{tab:pearson}. We found a very strong anticorrelation between $\sigma$ and  $L_\star$(z=0) (-0.90) and between $L_\star$(z=0) and $\phi_\star$(z=0) (-0.85). These classical strong correlations are due to the choice of the parametrisation of the LF. There are also very strong degeneracies between the evolution in density and in luminosity of the LF (-0.81 between 0 and the first break, -0.67 between the two breaks and -0.76 after the second break).\\

There are also some slight degeneracies between the calibration factors. The \textit{Spitzer} calibration parameters are correlated (0.68 between 24~$\mu$m and 70~$\mu$m, 0.73 between 24~$\mu$m and 160~$\mu$m, 0.62 between 70~$\mu$m and 160~$\mu$m). The other correlation implying a calibration factor are between -0.6 and 0.6.\\

\begin{figure*}
\centering
\includegraphics{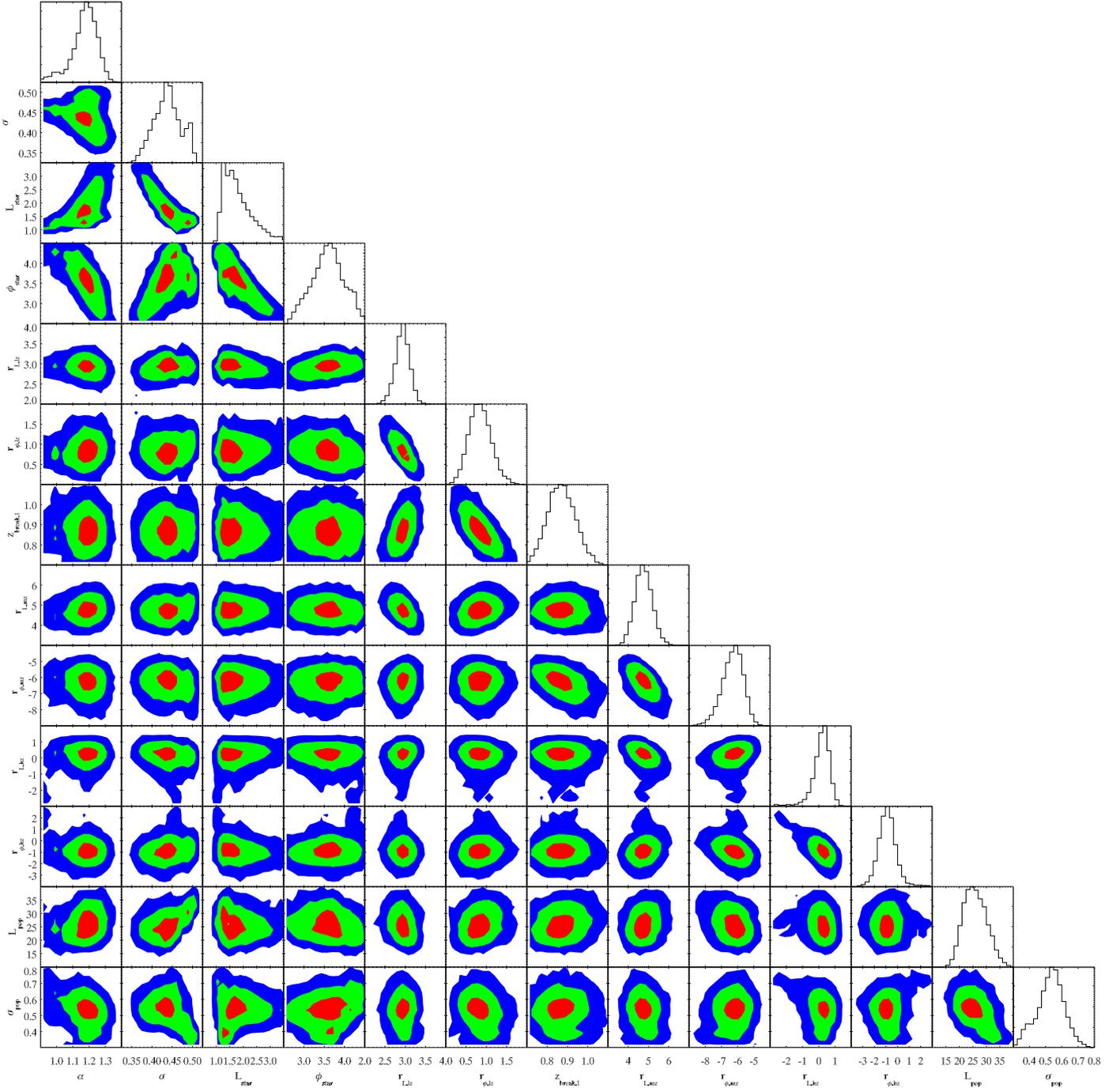}
\caption{\label{fig:contours} \textit{Diagonal plots}: marginalized probability distributions of each parameters deduced from the MCMC. \textit{Non-diagonal plots}: 1-$\sigma$ (\textit{red}), 2-$\sigma$ (\textit{green}) and 3-$\sigma$ (\textit{blue}) regions for each pair of parameters). \textit{From left to right:} $\alpha$, $\sigma$, L$_\star$, $\phi_\star$, $r_{L_\star,lz}$, $r_{phi_\star,lz}$, $z_{break,1}$, $r_{L_\star,mz}$, $r_{phi_\star,mz}$, $r_{L_\star,hz}$, $r_{phi_\star,hz}$, $L_{pop}$ and $\sigma_{pop}$ (c.f. Table \ref{tab:fitres}).}
\end{figure*}

The marginalized probability distributions of each parameter and the 1, 2, and 3-$\sigma$ confidence regions for each pair of parameters are plotted Fig. \ref{fig:contours}. Some distributions are not Gaussian. It thus justifies the use to use a MCMC algorithm.

\begin{table*}
\begin{tabular}{l|rrrr|rr|r|rr|rr|rr}
\hline
\hline
 &  $\alpha$ & $\sigma$ & $L_\star$(z=0) & $\phi_\star$ (z=0) & $r_{L_\star,lz}$ & $r_{phi_\star,lz}$ & $z_{break,1}$ & $r_{L_\star,mz}$ & $r_{phi_\star,mz}$ & $r_{L_\star,hz}$ & $r_{phi_\star,hz}$  & $L_{pop}$  & $\sigma_{pop}$ \\
\hline
$\alpha$ &  1.00 & -0.48 &  0.71 & -0.75 &  0.02 & -0.06 &  0.04 &  0.14 & -0.16 & -0.04 &  0.11 & -0.02 &  0.05 \\
$\sigma$ & -0.48 &  1.00 & -0.90 &  0.62 & -0.03 &  0.15 & -0.06 &  0.10 &  0.05 & -0.20 &  0.12 &  0.48 & -0.37 \\
$L_\star$(z=0)  &  0.71 & -0.90 &  1.00 & -0.85 & -0.14 & -0.03 &  0.07 &  0.00 & -0.11 &  0.11 & -0.04 & -0.19 &  0.20 \\
$\phi_\star$ (z=0) & -0.75 &  0.62 & -0.85 &  1.00 &  0.22 & -0.15 & -0.05 &  0.04 &  0.08 & -0.04 & -0.09 & -0.11 & -0.01 \\
\hline
$r_{L_\star,lz}$ &  0.02 & -0.03 & -0.14 &  0.22 &  1.00 & -0.81 &  0.51 & -0.44 &  0.10 &  0.14 & -0.12 & -0.27 &  0.13 \\
$r_{phi_\star,lz}$ & -0.06 &  0.15 & -0.03 & -0.15 & -0.81 &  1.00 & -0.78 &  0.18 &  0.07 & -0.08 &  0.13 &  0.18 & -0.17 \\
\hline
$z_{break,1}$ &  0.04 & -0.06 &  0.07 & -0.05 &  0.51 & -0.78 &  1.00 &  0.05 & -0.51 & -0.09 &  0.07 &  0.12 &  0.12 \\
\hline
$r_{L_\star,mz}$ &  0.14 &  0.10 &  0.00 &  0.04 & -0.44 &  0.18 &  0.05 &  1.00 & -0.67 & -0.43 &  0.29 &  0.05 & -0.09 \\
$r_{phi_\star,mz}$ & -0.16 &  0.05 & -0.11 &  0.08 &  0.10 &  0.07 & -0.51 & -0.67 &  1.00 &  0.35 & -0.41 & -0.04 & -0.07 \\
\hline
$r_{L_\star,hz}$ & -0.04 & -0.20 &  0.11 & -0.04 &  0.14 & -0.08 & -0.09 & -0.43 &  0.35 &  1.00 & -0.76 & -0.20 & -0.26 \\
$r_{phi_\star,hz}$ &  0.11 &  0.12 & -0.04 & -0.09 & -0.12 &  0.13 &  0.07 &  0.29 & -0.41 & -0.76 &  1.00 &  0.11 &  0.18 \\
\hline
$L_{pop}$ & -0.02 &  0.48 & -0.19 & -0.11 & -0.27 &  0.18 &  0.12 &  0.05 & -0.04 & -0.20 &  0.11 &  1.00 & -0.39 \\
$\sigma_{pop}$ &  0.05 & -0.37 &  0.20 & -0.01 &  0.13 & -0.17 &  0.12 & -0.09 & -0.07 & -0.26 &  0.18 & -0.39 &  1.00 \\
\hline
\end{tabular}
\caption{\label{tab:pearson} Pearson correlation matrix for our model. The part of the matrix concerning the calibration factors is not written to save space.}
\end{table*}

\begin{figure}
\centering
\includegraphics{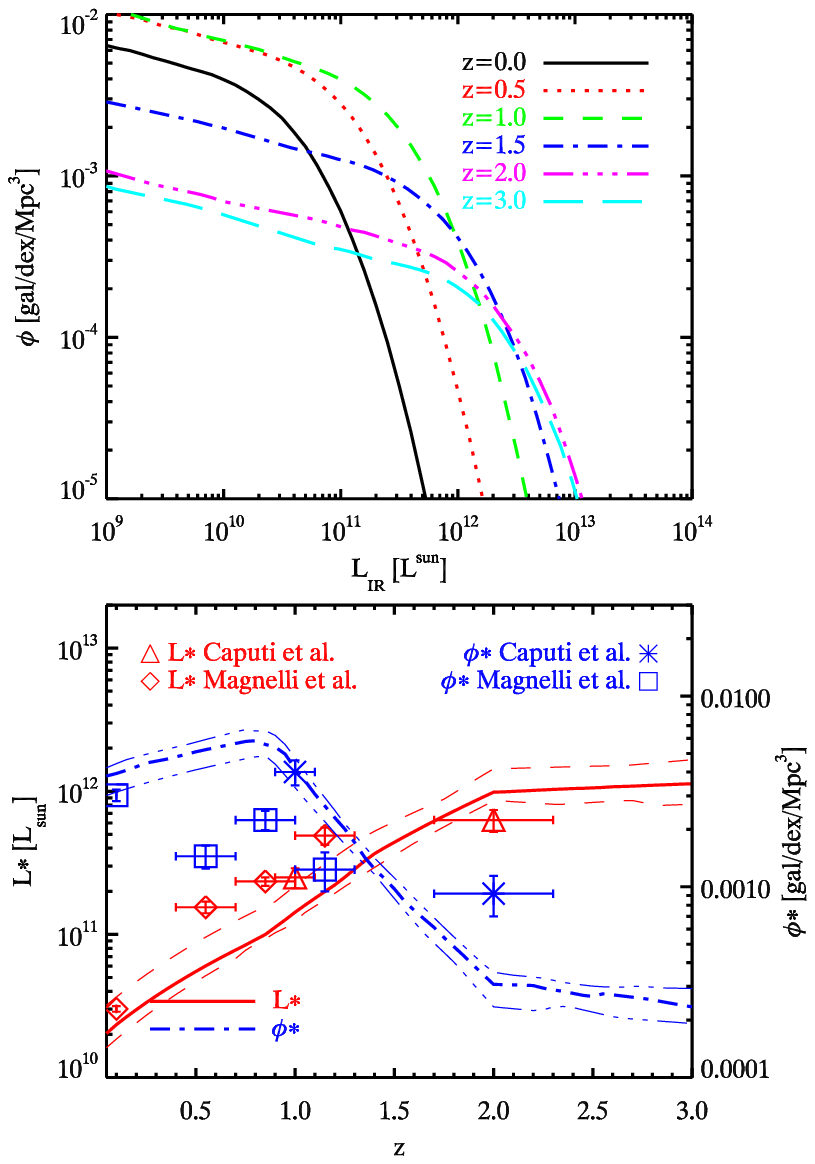}
\caption{\label{fig:lfevo} Evolution of the bolometric infrared luminosity function with the redshift. Upper panel: bolometric LF at z=0 (\textit{solid line}), z=0.5 (\textit{dot line}), z=1(\textit{dashed line}), z=1.5 (\textit{dot-dash line}), z=2 (\textit{3-dot-dash line}), z=3 (\textit{long dashed line}). Lower panel: Evolution of the $L_\star$ (\textit{red solid line}) and $\phi_\star$ (\textit{blue dot-dash line}) parameter as a function of redshift and 1-$\sigma$ confidence region. The measurement of $L_\star$ by \citet{Caputi2007} (\textit{triangles}) using 24~$\mu$m obervations and \citet{Magnelli2009} (\textit{diamonds})  using 70~$\mu$m observations are plotted in \textit{red}. The measurement of $\phi_\star$ by \citet{Caputi2007} (\textit{cross}) and \citet{Magnelli2009} (\textit{square}) are in \textit{blue}.}
\end{figure}

\section{Interpretation of the results}

\subsection{Evolution of the luminosity function}

Our model uses a very strong evolution of the bolometric infrared luminosity function to reproduce the infrared observations. The characteristic luminosity ($L_\star$) strongly decreases since z=2 to now. This parameter has been divided by about a factor 50 from z=2 to 0. The characteristic density ($\phi_\star$) increases strongly between z=2 and z=1 and slightly decreases between z=1 and now. At z$>$2, the model is compatible with no evolution in luminosity and a slight decrease of the density when the redshift increases. The evolution of these two parameters are plotted in Fig. \ref{fig:lfevo}.\\

We compared our results with \citet{Caputi2007} measurements performed from MIPS 24~$\mu$m observations and \citet{Magnelli2009} measurement obtained using MIPS 70~$\mu$m observations. These two works used a stacking analysis to measure the faintest points. The evolutions of $L_\star$ and $\phi_\star$ only marginally agrees with these two works. Nevertheless, they use different fixed values of $\sigma$ and $\alpha$ and an extrapolation from the monochromatic luminosity to $L_{IR}$. These choices could imply some biases. We found as \citet{Caputi2007} a strong negative evolution in density between z$\sim$1 and z$\sim$2. They found an evolution in (1+z)$^{-3.9\pm1.0}$ and we found (1+z)$^{-6.2\pm0.5}$. Nevertheless, our value is probably biased by our non-smooth parametrization. This evolution is discussed in details by \citet{Caputi2007}.\\

\citet{Reddy2008} claimed that $\alpha \sim 1.6$ at z$>$2. But, we do not need an evolution of $\alpha$ and $\sigma$ to reproduce the observations. Nevertheless, the infrared measurements are not sufficiently deep to constraint accurately an evolution of $\alpha$.

\subsection{Evolution of the dust-obscured star formation rate}

\begin{figure}
\centering
\includegraphics{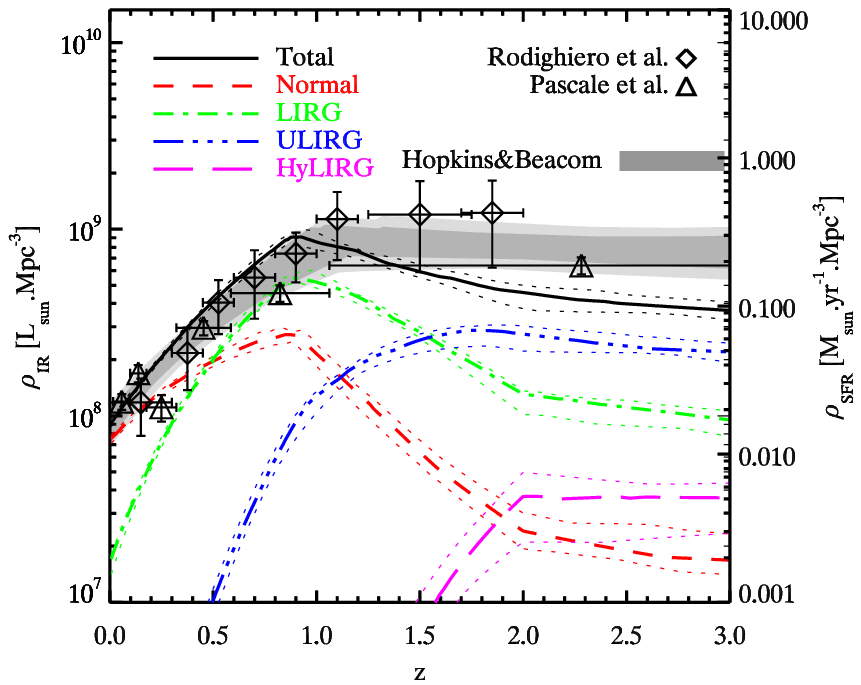}
\caption{\label{fig:rhoir} Evolution of the bolometric infrared luminosity density (\textit{black solid line}) as a function of the redshift. The contribution of normal galaxies ($L_{IR}<10^{11} \, L_\odot$), LIRG($10^{11}<L_{IR}<10^{12} \, L_\odot$), ULIRG($10^{12}<L_{IR}<10^{13} \, L_\odot$), HyLIRG($L_{IR}>10^{13} \, L_\odot$) are plotted with \textit{short-dahsed}, \textit{dot-dash}, \textit{three-dot-dash}, \textit{long-dashed line} respectively. The measurements of \citet{Rodighiero2009} using the MIPS 24~$\mu$m data are plotted with \textit{diamonds} and \citet{Pascale2009} ones using a BLAST stacking analysis with \textit{triangles}. The \citet{Hopkins2006} fit of the optical and infrared measurement is plotted with a \textit{dark grey area} (1-$\sigma$) and a \textit{light grey area} (3-$\sigma$).}
\end{figure}

The bolometric infrared luminosity density ($\rho_{IR}$) can be deduced from the bolometric infrared LF. Our local value of $\rho_{IR}$ ((1.05$\pm$0.05)$\times$10$^8$~L$_\odot$.Mpc$^{-3}$) agrees with \citet{Vaccari2010} measurements  (1.31$_{-0.21}^{+0.24}\times$10$^8$~L$_\odot$.Mpc$^{-3}$). We also agree well with measurements at higher redshift (\citet{Rodighiero2009} and \citet{Pascale2009} (see Fig. \ref{fig:rhoir}). $\rho_{IR}$ can be converted into star formation rate density (SFRD) using the conversion factor 1.7$\times$10$^{-10}\,M_\odot.yr^{-1}.L_\odot^{-1}$ \citep{Kennicutt1998}. The SFRD derived from our model agrees rather well with the \citet{Hopkins2006} fit of the optical and infrared measurements.\\

We also determined the contribution of the different ranges of luminosity (normal: $L_{IR}<10^{11}\,L_\odot$, LIRG: $10^{11}<L_{IR}<10^{12} \, L_\odot$, ULIRG:$10^{12}<L_{IR}<10^{13} \, L_\odot$, HyLIRG: $L_{IR}>10^{13} \, L_\odot$). Between z=0 and 0.5, the infrared luminosity density is dominated by normal galaxies ($L_{IR}<10^{11} \, L_\odot$). Their contribution decreases slowly with redshift due to the evolution of the LF seen in Fig. \ref{fig:lfevo}. Between  z=0.5 and 1.5, the infrared output is dominated by the LIRG. At higher redshift, it is dominated by ULIRGs. The HyLIRGs never dominate and account for some percent at high redshift. A physical cutoff at very high luminosity thus would not change strongly the infrared density evolution.\\

Following our model, the number of very bright objects ($>10^{12.5}~L_\odot$) is maximal around z=2 (see Fig. \ref{fig:lfevo}). These objects could be very massive galaxies observed during their formation in the most massive dark matter halos. Among other analysis, the study of the spatial distribution of the galaxies will help to confirm or infirm this scenario \citep{Penin2010}.\\

Around z=1, the number of very bright objects is lower than at higher redshift, but the number of LIRGs is about one order of magnitude larger. From z=1 to now, the infrared output has decreased by about one order of magnitude. Our model makes only a description of this evolution and we need physical models to understand why, contrary to nowadays, the star formation at high redshift is dominated by few very-quickly-star-forming galaxies, when the associated dark matter halos grew by hierarchical merging \citep{Cole2000,Lazoni2005}. We also need an explanation of the decrease of the star formation since z=1. The main candidates the feedback of AGNs and starbursts (e.g. \citet{Baugh2006}) and/or the lack of gas.

\subsection{CIB SED}

\begin{figure}
\centering
\includegraphics{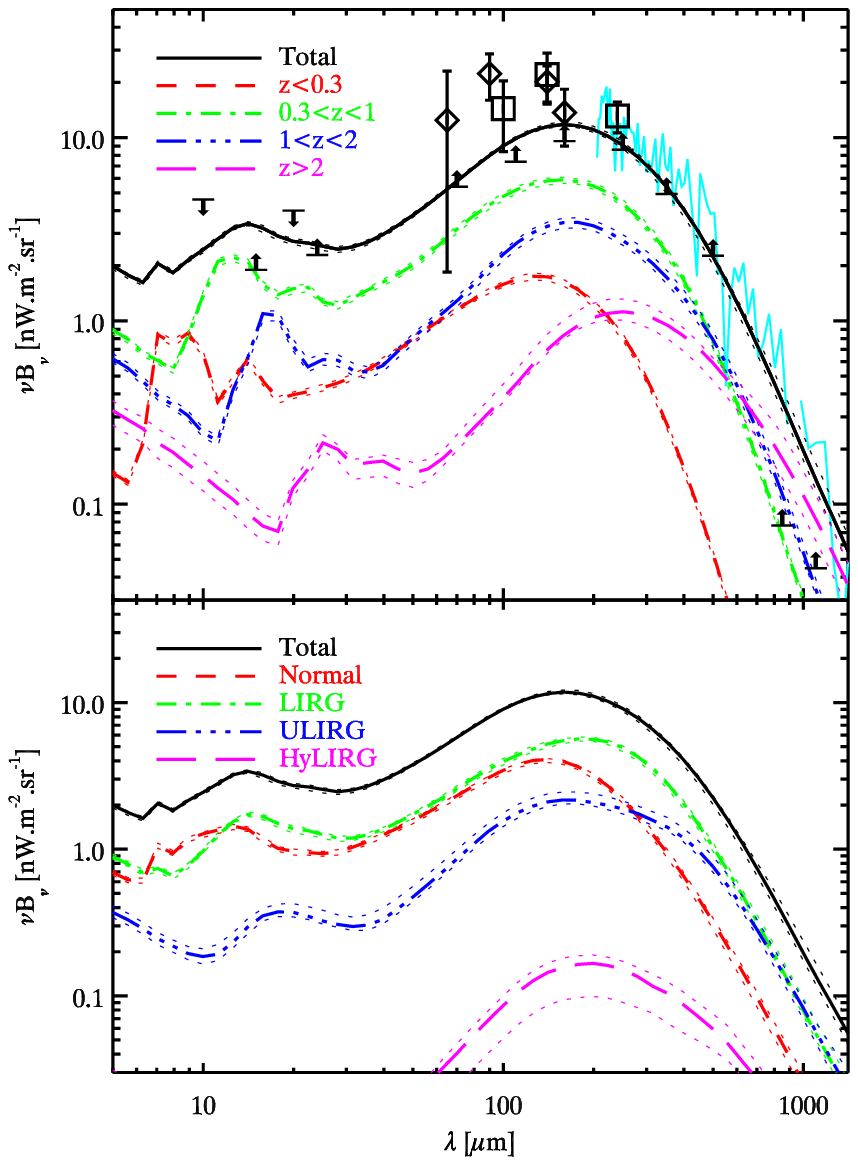}
\caption{\label{fig:cib} Upper panel: contribution to the CIB per redshift slice. \textit{Black solid line}: CIB spectrum predicted by the model. \textit{Red short-dashed line}: Contribution of the galaxies between z=0 and 0.3. \textit{Green dot-dash line}: Same thing between z=0.3 and 1. \textit{Blue three dot-dash line}: same thing between z=1 and 2. \textit{Purple long-dashed line}: Contribution of the galaxies at redshift larger than 2. \textit{Black arrows}: Lower limits coming from the number counts at 15~$\mu$m \citep{Hopwood2010} and 24~$\mu$m \citep{Bethermin2010a} and the stacking analysis at 70~$\mu$m \citep{Bethermin2010a}, 100~$\mu$m, 160~$\mu$m \citep{Berta2010}, 250~$\mu$m, 350~$\mu$m, 500~$\mu$m \citep{Marsden2009}, 850~$\mu$m \citep{Greve2009} and 1.1~mm \citep{Scott2010} and upper limits coming from absorption of the TeV photons of \citet{Stecker1997} at 20~$\mu$m and \citet{Renault2001} between 5~$\mu$m and 15~$\mu$m. \textit{Black diamonds}: \citet{Matsuura2010} absolute measurements with Akari. \textit{Black square}: \citet{Lagache2000} absolute measurements with DIRBE/WHAM. \textit{Cyan line}: \citet{Lagache2000} FIRAS measurement. \newline Lower panel: Contribution of to the CIB of the normal galaxies (\textit{red short-dashed line}), LIRGs ({green dot-dash line}), ULIRGs (\textit{blue three dot-dash line}) and HyLIRG (\textit{purple long-dashed line}) and all the galaxies (\textit{black solid line}).}
\end{figure}

The value of the CIB at different wavelengths predicted by the model is given in Table \ref{tab:ciblevel}. We found a CIB integrated value (over the 8-1000~$\mu$m range) of 23.7$\pm$0.9~nW.m$^{-2}$.sr$^{-1}$. It agrees with the 24-27.5~nW.m$^{-2}$.sr$^{-1}$ range of \citet{Dole2006}.\\

We compared the CIB spectrum found with our model with the measurements (see Fig. \ref{fig:cib}). Our model is always higher than the lower limit given by the stacking. The \citet{Marsden2009} limits are very stringent. Nevertheless, they could be overestimated due to the contamination due to clustering \citep{Bavouzet2008,Fernandez_Conde2010,Bethermin2010b}. Our model is compatible with the upper limit given by the absorption of the TeV photons by photon-photon interaction with the CIB (see Sect. \ref{sect:tev}). We globally agree with the DIRBE/WHAM \citep{Lagache2000} and Akari \citep{Matsuura2010} absolute measurement, excepting at 90~$\mu$m (Akari) and 100~$\mu$m (DIRBE/WHAM) where the measurements are significantly higher than our model. These measurements need an accurate subtraction of the zodiacal light and of the galactic emissions and an accurate inter-calibration between DIRBE and FIRAS. Indeed, a bad removal of the zodiacal light explains this disagreement \citep{Dole2006}. At larger wavelengths, we very well agree with the FIRAS absolute measurements of \citet{Lagache2000}.\\

We separated the contribution of the infrared galaxies to the CIB in 4 redshift slices, each slice corresponding to about a quarter of the age of the Universe (Fig. \ref{fig:cib}). Between 8 and 30~$\mu$m, we can see a shaky behavior of each slice due to the PAH emission bands. The total is smoother. The $0<z<0.3$ dominates the spectrum only near 8~$\mu$m due to the strong PAH emission at this rest-frame wavelength. This slice, where the infrared luminosity density is the lowest, has a minor contribution at the other wavelengths. The $0.3<z<1$ slice dominates the spectrum between 10 and 350~$\mu$m. The sub-mm and mm wavelengths are dominated by the sources lying at higher redshift ($z>2$, see \citet{Lagache2005}). It is due to redshift effects that shift the peak of emission around 80~$\mu$m rest-frame to the sub-mm. The mean redshift of the contribution to the CIB is written in Table \ref{tab:ciblevel} and computed with
\begin{equation}
<z> = \frac{\int_{0}^{\infty} z \frac{dB_{\nu}}{dz} dz}{\int_{0}^{\infty} \frac{dB_{\nu}}{dz} dz}\\
\end{equation}

\begin{table}
\centering
\begin{tabular}{lrrr}
\hline
\hline
wavelength & CIB & CIB & $<$z$>$\\
$\mu$m & nW.m$^{-2}$.sr$^{-1}$ & MJy.sr$^{-1}$ & \\
\hline
15 &      3.294$^{+     0.105}_{-     0.128}$ &      0.016$^{+     0.001}_{-     0.001}$ &      0.820$^{+     0.026}_{-     0.018}$ \\
24 &      2.596$^{+     0.076}_{-     0.139}$ &      0.021$^{+     0.001}_{-     0.001}$ &      0.894$^{+     0.025}_{-     0.029}$ \\
70 &      5.777$^{+     0.165}_{-     0.067}$ &      0.135$^{+     0.004}_{-     0.002}$ &      0.773$^{+     0.022}_{-     0.021}$ \\
100 &      9.014$^{+     0.231}_{-     0.125}$ &      0.300$^{+     0.008}_{-     0.004}$ &      0.829$^{+     0.023}_{-     0.024}$ \\
160 &     11.771$^{+     0.382}_{-     0.318}$ &      0.628$^{+     0.020}_{-     0.017}$ &      0.947$^{+     0.032}_{-     0.019}$ \\
250 &      9.100$^{+     0.395}_{-     0.382}$ &      0.758$^{+     0.033}_{-     0.032}$ &      1.124$^{+     0.053}_{-     0.021}$ \\
350 &      5.406$^{+     0.190}_{-     0.417}$ &      0.631$^{+     0.022}_{-     0.049}$ &      1.335$^{+     0.075}_{-     0.060}$ \\
500 &      2.237$^{+     0.077}_{-     0.217}$ &      0.373$^{+     0.013}_{-     0.036}$ &      1.680$^{+     0.124}_{-     0.122}$ \\
850 &      0.374$^{+     0.057}_{-     0.042}$ &      0.106$^{+     0.016}_{-     0.012}$ &      2.444$^{+     0.292}_{-     0.192}$ \\
1100 &      0.139$^{+     0.031}_{-     0.017}$ &      0.051$^{+     0.011}_{-     0.006}$ &      2.833$^{+     0.341}_{-     0.201}$ \\
\hline
\end{tabular}
\caption{\label{tab:ciblevel} Surface brightness of the CIB and mean redshift $<$z$>$ of the contribution to the CIB at different wavelengths as predicted by the model.}
\end{table}

We also separate the contribution of the different infrared luminosity classes. The normal galaxies and LIRGs dominate the background up to 250~$\mu$m. It is compatible with the fact that these populations are dominant at low redshift. At larger wavelengths, the redshift effects tend to select high redshift sources; LIRGs and ULIRGs are responsible for about half of the CIB each. The HyLIRG have only a small contribution ($<10\%$) including in the mm range (Fig. \ref{fig:cib}, bottom).

\section{Predictions}

\subsection{Confusion limit}

\label{section:confusion}

\begin{figure}
\centering
\includegraphics{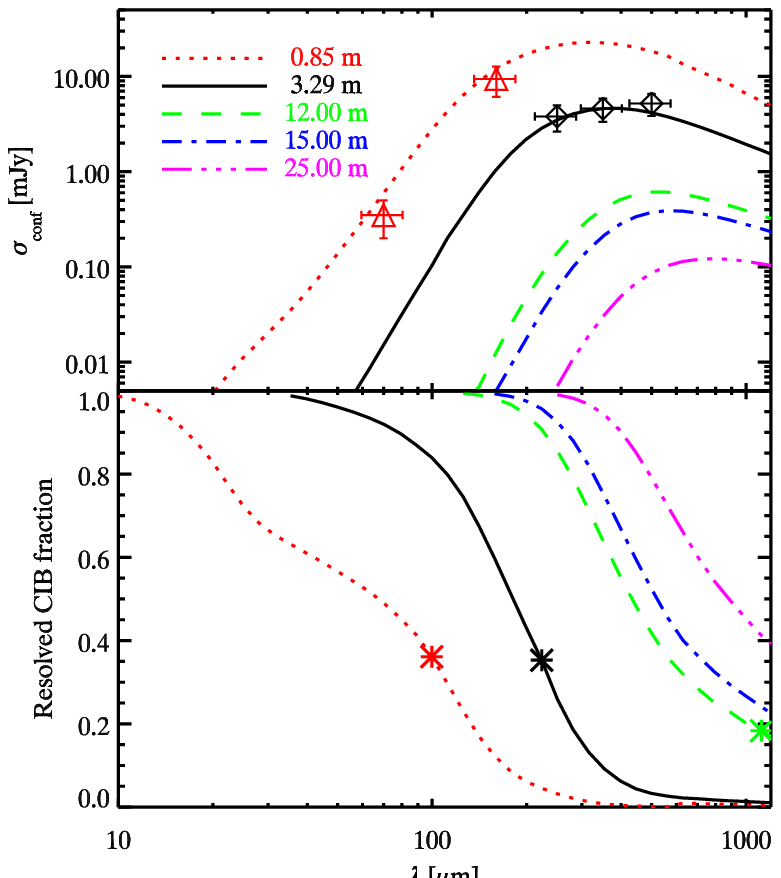}
\caption{\label{fig:confusion} Upper panel: 1-$\sigma$ confusion noise as a function of the wavelength for different telescope diameters. We use the confusion noise given by the P(D) method (see Sect. \ref{section:confusion}) for this plot. \textit{Red triangles}: \citet{Frayer2009} \textit{Spitzer}/MIPS confusion measurements. \textit{Black diamonds}: \citet{Nguyen2010}: \textit{Herschel}/SPIRE confusion noise measurements (5-$\sigma_{conf}$ cut). \newline Lower panel: Resolved fraction of the CIB by sources brighter than 5-$\sigma$ confusion noise (fluctuations) and the source density limit. \newline Both panel: \textit{Red dotted line}: Telescope with a diameter of 0.85~m like Spitzer. \textit{Black solid line}: 3.29~m telescope like \textit{Herschel}. \textit{Green dashed line}: 12~m telescope like Atacama pathfinder experiment (APEX). \textit{Blue dot-dashed line}: 15~m telescope like the CSO. \textit{Purple three dot-dashed line}: 25~m like the CCAT project. \textit{Asterisks}: Transition between the source density limitation (short wavelengths) and the fluctuation limitation (long wavelengths).}
\end{figure}

The confusion limit can be defined in several ways. The radioastronomers use classically a source density criteria, where the confusion limit is the flux cut for which a critical density of source is reached. The choice of this critical density is not trivial. We follow the approach of \citet{Dole2003}. The source density limit $N_{SDC}$ is reached when there is a probability P to have an other source in a k~$\theta_{FWHM}$ radius (where $\theta_{FWHM}$ is the full width at half maximum of the beam profile). \citet{Dole2003} show that
\begin{equation}
N_{SDC} = -\frac{log(1-P)}{\pi k^2 \theta_{FWHM}^2}
\end{equation}
We chose P = 0.1 and k = 0.8 following \citet{Dole2003}.\\

This source density criterion does not take into account the contributions of the sources fainter than the flux limit. We made also an estimate of the photometric confusion noise based on the P(D) analysis (see Sect. \ref{sect:pd}). The P(D) distribution in absence of instrumental noise is not gaussian and have a large tail at bright flux. Thus the standard deviation is not a good estimator of the confusion noise. We chose to compute the interquartile interval of the P(D) divided by 1.349. With this definition, the value of the confusion noise is exactly $\sigma$ in the Gaussian case, and we are less sensitive to the bright outliers.\\

These two estimators can be computed from the counts predicted by our model. We assume that the sources are point-like. The confusion noise found for large telescope at short wavelength ($<$8~$\mu$m for a 0.85~m-diameter telescope like \textit{Spitzer} and $<$35~$\mu$m for a 3.29~m-diameter telescope like \textit{Herschel}) are thus underestimated. For this reason, we do not estimate the confusion levels for beam smaller than 2 arcsec.\\

The Fig. \ref{fig:confusion} (upper panel) represents the confusion noise. It agrees with the confusion noise measured by \citet{Frayer2009} and \citet{Nguyen2010} with \textit{Spitzer}/MIPS and \textit{Herschel}/SPIRE. \citet{Weiss2009} estimate that the confusion noise for a APEX/LABOCA map smoothed by the beam is 0.9~mJy/beam. We find 0.6~mJy/beam with the P(D) approach.\\

We also compute the resolved fraction of the CIB by sources brighter than the confusion limit of \citet{Dole2003} (source density criterion) and the 5-$\sigma_{conf}$ given by the P(D). Fig. \ref{fig:confusion} (lower panel) and Table \ref{tab:0085}, \ref{tab:0329}, \ref{tab:1200}, \ref{tab:1500} and \ref{tab:2500} summarize the results. The transition in the confusion regime between the source density limitation (short wavelengths) and the fluctuation limitation (long wavelengths) happens at 100~$\mu$m for \textit{Spitzer}, 220~$\mu$m for \textit{Herschel} and 1120~$\mu$m for the CSO (asterisks in the lower panel of Fig. \ref{fig:confusion}). For larger antennas below 1.2~mm, the confusion is mainly due to the source density.\\

According to these results, at the confusion limit, \textit{Herschel} can resolve 92\%, 84\%, 60\%, 25.9\%, 9.2\% and 3.3\% of the CIB at 70~$\mu$m, 100~$\mu$m, 160~$\mu$m, 250~$\mu$m, 350~$\mu$m and 500~$\mu$m respectively. Nevertheless, due to the blackbody emission of the telescope (about 60~K), very long integration times are needed to reach the confusion limit at short wavelengths. The confusion limit in PACS will be reach only in the ultra-deep region of the H-GOODS survey. The confusion limit will probably be never reached at 70~$\mu$m. A telescope with the same size as \textit{Herschel} and a cold (5K) mirror, like SPICA, could resolve almost all the CIB from the mid-infrared to 100~$\mu$m. A 25~m single-dish sub-mm telescope like the Cornell Caltech Atacama telescope (CCAT) project would be able to resolve more than 80\% of the CIB up to 500~$\mu$m.

\subsection{High energy opacity}

\label{sect:tev}

The CIB photons can interact with TeV photons. The cross section between a $E_\gamma$ rest-frame high-energy photon and an infrared photon with a observer-frame wavelength $\lambda_{IR}$ interacting at a redshift $z$ with an angle $\theta$ (and $\mu = cos(\theta)$) is \citep{Heitler1954,Jauch1976}
\begin{eqnarray}
\sigma_{\gamma \gamma}(E_\gamma,\lambda_{IR},\mu,z) = H\left (1-\frac{\epsilon_{th}}{\epsilon} \right ) \frac{3 \sigma_T}{16} (1-\beta^2) \times \\
 \left [ 2\beta(\beta^2-2)+(3-\beta^4)ln\left( \frac{1+\beta}{1-\beta} \right) \right ]
\end{eqnarray}
with
\begin{equation}
\beta = \sqrt{1-\frac{\epsilon_{th}}{\epsilon}},
\end{equation}
\begin{equation}
\epsilon_{th}(E_\gamma,\mu,z) = \frac{2(m_e c^2)^2}{E_\gamma (1-\mu)(1+z)},
\end{equation}
\begin{equation}
\epsilon(\lambda_{IR},z) = \frac{hc(1+z)}{\lambda_{IR}},
\end{equation}
where $\sigma_T$ is the Thompson cross section ($6.65\times10^{-29}$~m$^{2}$), $m_e$ the mass of the electron and H the Heaviside step function (H(x)=1 if x$>$0 and 0 else).\\

\begin{figure}
\centering
\includegraphics{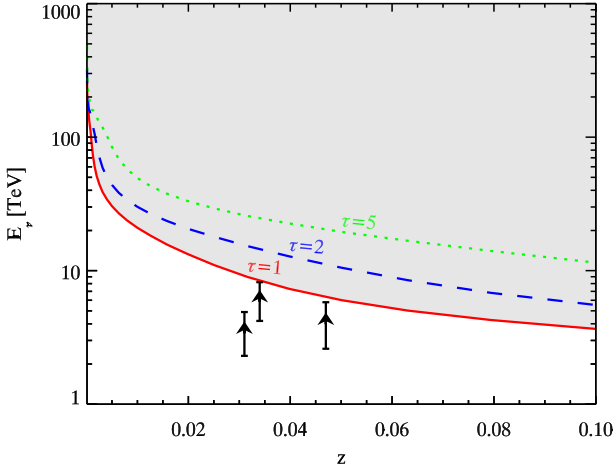}
\caption{\label{fig:tevopacity} Fazio-Stecker relation: energy at which the opacity reach a given $\tau$ as a function of redshift. This plot is done for $\tau$ = 1 (\textit{red solid line}), 2 (\textit{blue dashed line}) and 5 (\textit{green dotted line}). The data points are the cutoff energy of Mkn 501 \citep{Aharonian1999}, Mkn 421 \citep{Aharonian2002} and Lac 1ES 1959+650 \citep{Aharonian2003}.}
\end{figure}

The optical depth $\tau(E_\gamma,z_s)$ for a photon observed at energy $E_\gamma$ and emitted at a redshift $z_s$ can be easily computed \citep{Dwek2005,Younger2010,Dominguez2010} with
\begin{eqnarray}
\tau(E_\gamma,z_s) =  \int_0^{z_s}  dz \frac{D_H}{\sqrt{\Omega_\Lambda + (1+z)^3 \Omega_m}} \\
 \int_{-1}^1 d\mu \frac{1-\mu}{2} \int_{5 \,\mu m}^\infty d\lambda_{IR}  n_{\lambda_{IR}}(\lambda_{IR},z) (1+z)^2 \sigma_{\gamma \gamma}(E_\gamma,\lambda_{IR},\mu,z)
\end{eqnarray}
where $n_{\lambda_{IR}}(\lambda_{IR},z)$ is the comoving number density of photons emitted at a redshift greater than z between $\lambda_{IR}$ and $\lambda_{IR}+d\lambda_{IR}$. The 5~$\mu$m cut corresponds to the limit of the validity of our model. The number density of photons is computed with 
\begin{equation}
n_{\lambda_{IR}}(\lambda_{IR},z) = \frac{4 \pi}{h c \lambda_{IR}} (B_{\nu,CIB}+B_{\nu,CMB})
\end{equation}
where $B_{\nu,CIB}$ is the CIB given by our model and $B_{\nu,CMB}$ is the brightness of a blackbody at 2.725K corresponding to the cosmic microwave background \citep{Fixsen2009}. Our predicted opacities do not take into account the absorption by the cosmic optical background photons (COB, $\lambda<5$~$\mu$m). \citet{Younger2010} showed that the contribution of the COB to the opacity is negligible for energies larger than 5~TeV.\\

We can determine up to which redshift the opacity stays lower than 1. We can thus define an horizon as a function of the energy, called Fazio-Stecker relation. We can see in Fig. \ref{fig:tevopacity} that the observed energy cutoff of low-redshift blazars (Mkn 501 \citep{Aharonian1999}, Mkn 421 \citep{Aharonian2002} and BL Lac 1ES 1959+650 \citep{Aharonian2003}) is compatible  with this relation.

\subsection{Effect of the strong lensing on the number counts}

\label{section:stronglensing}

\begin{figure}
\centering
\includegraphics{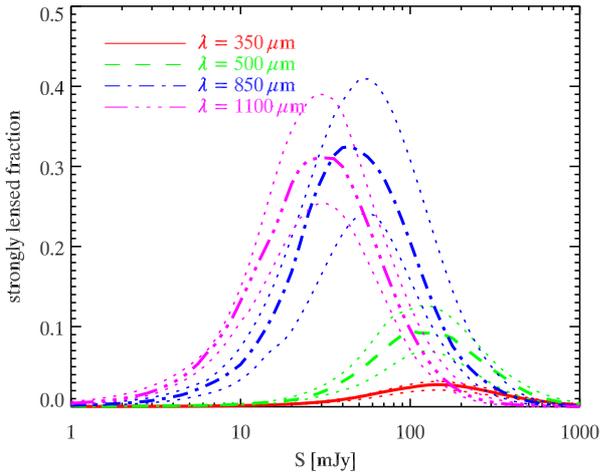}
\caption{\label{fig:lensed_fraction} Fraction of strongly lensed sources (magnification larger than 2) as a function of the flux at 350~$\mu$m (\textit{red solid line}), at 500~$\mu$m (\textit{green dashed line}), at 850~$\mu$m (\textit{blue dot-dashed line}) and at 1.1~mm (\textit{purple three-dot-dashed line}). The \textit{dotted lines} represent the 1-$\sigma$ confidence area of our model.}
\end{figure}

The strongly-lensed fraction is the ratio between the counts of lensed sources and the total observed counts. Because the slope of the counts varies a lot with the flux and wavelength, this fraction depends on the flux and the wavelength (see Fig. \ref{fig:lensed_fraction}). The strongly lensed fraction is always smaller than 2\% below 250~$\mu$m and is thus negligible. At larger wavelengths, we predict a maximum of the strongly lensed fraction near 100~mJy. At 500~$\mu$m, about 15~\% of the sources brighter than 100~mJy are lensed. This fraction increases to 40\% near 1~mm.\\

Our results can be compared with ones of \citet{Negrello2007} model. The two model predict that the lensed fraction as a function of the flux is a bump around 100~mJy. But, the amplitude of this bump predicted by the two models is significantly different. For instance, the maximum of the lensed fraction at 500~$\mu$m is 15\% for our model and 50\% for the \citet{Negrello2007} model. The slope between 10 and 100~mJy  is steeper in \citet{Negrello2007} model than in ours and is incompatible with the measurements \citep{Clements2010,Oliver2010,Glenn2010}. The steeper the slope, the larger the lensed fraction. This explains why the \citet{Negrello2007} model predicts larger lensed fraction than ours. The probability for a source to be lensed increases with its redshift. Differences in the redshift distributions of the models could also explain some differences in the lensed fraction.\\

The Fig. \ref{fig:spt_counts} shows the respective contribution of the lensed and non-lensed sources to the SPT counts of dusty sources without IRAS 60~$\mu$m counterparts at 1.38~mm \citep{Vieira2009}. According to the model, these counts are dominated by strongly-lensed sources above 15~mJy. These bright sources are thus very good candidates of strongly-lensed sub-mm galaxy.\\

\begin{figure}
\centering
\includegraphics{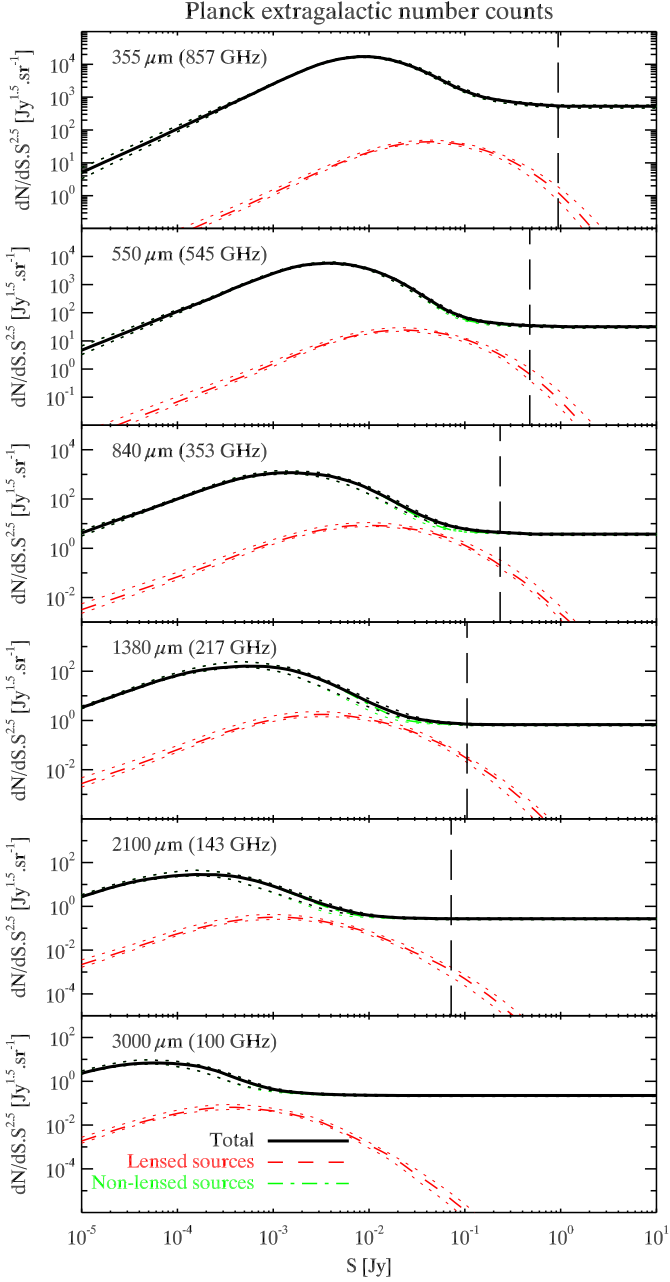}
\caption{\label{fig:planck_counts} Differential number counts in the \textit{Planck} bands. These counts takes only into account the individual star forming galaxies. \textit{Black solid line}: Total contribution. \textit{Green dot-dashed line}: Contribution of the non-lensed sources. \textit{Red dashed line}: Contribution of the strongly-lensed sources. \textit{Dotted lines} 1-$\sigma$ contours. \textit{Vertical long-dahsed line}: 5-$\sigma$ limits (confusion+instrumental) of \citet{FernandezConde2008} for a bias of 1.5.}
\end{figure}

We made predictions on the contribution of the strongly-lensed sources to the \textit{Planck} number counts (see Fig. \ref{fig:planck_counts}). We use the \citet{FernandezConde2008} 5-$\sigma$ limits, because they take into account the effect of the clustering on the confusion noise. This effect is non-negligible due to the large beam of \textit{Planck}. We found that the contribution of the lensed sources to the \textit{Planck} counts is negligible in all the bands (a maximum of 0.47 galaxies.sr$^{-1}$ at 550~$\mu$m). At high redshift, \textit{Planck} will probably detect more small structures like proto-clusters, than individual galaxies. \textit{Planck} is thus not the best survey to find lensing candidates. Sub-mm surveys with a sensitivity near 100~mJy are more efficient. For instance, the \textit{Herschel}-ATLAS survey should found 153$\pm$26 and 411$\pm$24 lensed sources with $S_{500}>$50~mJy and  $S_{350}>$50~mJy, respectively, on 600 deg$^2$.

\section{Discussion}

\subsection{Comparison with other backwards evolution models}

The evolution of the infrared luminosity density predicted by our model can be compared with the prediction of the recent backwards evolution models. We find, like \citet{Franceschini2009}, a strong increase of $\rho_{IR}$ from z=0 to z=1, a break around z=1, and a decrease at larger redshift. On the contrary, the \citet{Valiante2009} and \citet{Le_Borgne2009} models predict a maximum of infrared luminosity density around z=2.\\

As \citet{Le_Borgne2009} and \citet{Franceschini2009}, we found that LIRGs dominate infrared luminosity density around z=1 and that ULIRG dominate at redshift larger than 1.5. We also found as \citet{Le_Borgne2009} that normal galaxies dominates only up to z$\sim$0.5.\\

The \citet{Valiante2009} and our model use a similar parametrization of the evolution the LF which can be compared. Both models found a very strong evolution in luminosity up to z=2 ($(1+z)^{3.4}$ for the \citet{Valiante2009} model; $(1+z)^{2.9\pm0.1}$ from z=0 to 0.87$\pm$0.05 and $(1+z)^{4.7\pm0.3}$ from z= 0.87$\pm$0.05 to 2 for our model). At larger redshift, our model is compatible with no evolution and the \citet{Valiante2009} model predicts a slight decrease in $(1+z)^{-1}$. Concerning the evolution in density, both models predicts an increase from z=0 to z$\approx$1 (in $(1+z)^2$ for the \citet{Valiante2009} model and in $(1+z)^{0.8\pm0.2}$ for our model) and a decrease at larger redshift ($(1+z)^{-1.5}$ for the \citet{Valiante2009} model, $(1+z)^{-6.2\pm0.5}$ between z=0.87$\pm$0.5 and z=2 and $(1+z)^{-0.9\pm0.7}$ at z$>$2 for our model). These two models thus agree on the global shape of the evolution of the LF, but disagree on the values of the coefficient driving it. There is especially a large difference on the evolution density between z$\sim$ and z$\sim$2. This difference could be explained by different positions of the breaks. Nevertheless, the uncertainties on the \citet{Valiante2009} model are not estimated. It is thus hard to conclude.\\

\citet{Valiante2009} and \citet{Franceschini2009} used AGNs to reproduce the infrared observations. \citet{Valiante2009} also used a temperature dispersion of the galaxies. Our model reproduce the same observations using neither AGNs nor temperature dispersion. This show that the AGN contribution and the temperature scatter cannot be accurately constraint with this type of modeling.\\

\subsection{Discriminating the models: smoking guns observations?}

Although they use different galaxy populations and evolutions, the backwards evolution models reproduce the number counts from the mid-IR to the mm domain in a reasonably good way. It is thus important to find new observables to discriminate between models.\\

The comparison with the sub-mm redshift distributions of the bright sources is a rather simple, but very discriminant observations. For instance, the Fig. \ref{fig:nz} shows significant difference of the sub-mm redshift distributions predicted by the different models. The \citet{Chapin2010} measurements performed on one small field with a cut at high flux is not sufficient to conclude. \textit{Herschel} will help to increase the accuracy of the measured redshift distributions and to estimate the cosmic variance on them. These constraints will be crucial for the next generation of models.\\

\begin{figure}
\centering
\includegraphics{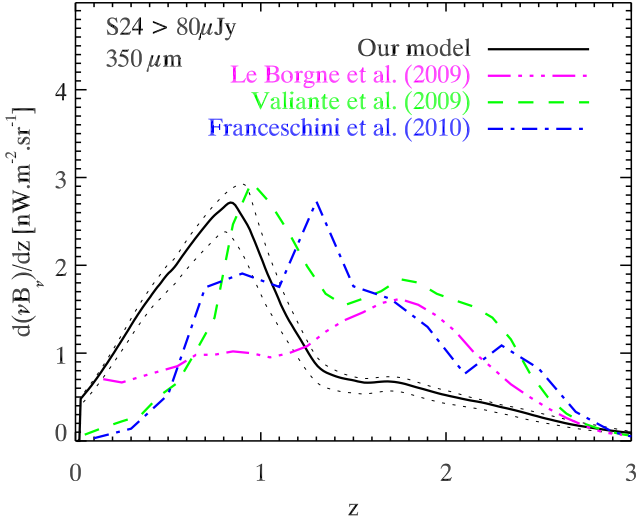}
\caption{\label{fig:dbdz_submm} Differential contribution of the S$_{24}>80$~$\mu$Jy sources to the CIB as a function of the redshift at 350~$\mu$m. \textit{Black solid line}: Our model (1-$\sigma$ limit in \textit{black dotted line}). \textit{Purple three dot-dashed line}: \citet{Le_Borgne2009} model. \textit{Green dashed line}: \citet{Valiante2009} model. \textit{Blue dot-dashed line}: \citet{Franceschini2009} model.}
\end{figure}

\citet{Jauzac2010} showed that the redshift distribution of the contribution of the 24 microns sources to CIB at 70 and 160~$\mu$m  ($d(\nu B_{\nu})/dz$) is also a very discriminant constraint. The Fig. \ref{fig:dbdz_submm} shows the $d(\nu B_{\nu})/dz$ at 350~$\mu$m. The different models make totally incompatible predictions in the sub-mm. An accurate measurement of $d(\nu B_{\nu})/dz$ will be thus crucial for the future models.\\

\subsection{Limits of our model}

Our model is a useful tool to make a first interpretation of the observations from the mid-infrared to the mm domain. Nevertheless it is biased by some structural choice in its construction.\\

The choice of the parameters biases the results. For example, we have chosen the minimal number of parameters to reproduce the counts. If we would used more breaks in the evolution in density and in luminosity, the evolutions with the redshift would be smoother and the errors on the predictions would be different. Our errors are just the statistical errors due to the determination of the parameter of a given model using the data. It does not include the uncertainty on our hypothesis on the evolution (like $\alpha$ fixed) and the biases due to our choice of parameters (evolution in $(1+z)^r$ with breaks). For instance, the strong decrease in density between z$\sim$0.9 and z=2 is probably an artifact due to our choice of parametrization. In addition, our model of lensing is very simple and should be updated in the future. Nevertheless, the contribution of the lensing in the fitted data is low and the bias is thus negligible.\\

The backwards evolution models gives a very limited interpretation of the data. They are only a description of the evolution of the statistical properties of the infrared galaxies. The physical processes explaining the strong evolution of these objects are ignored. A more complex physical approach is thus necessary to deeply understand the history of the infrared galaxies. Nevertheless, our model is very useful to make a rapid interpretation of new observations and predictions for the future missions.\\

\subsection{Perspectives}

Our model fit the current data with rather simple hypotheses. Nevertheless, the increasing accuracy of the infrared observations will probably help us to refine it. Lots of updates will be possible when we will need a more complex model.\\

The $\alpha$ and $\sigma$ parameters are fixed, but we can imagine an evolution of the shape of the LF with the redshift. A Fisher matrix analysis shows that the evolution of $\alpha$ at high redshift cannot be constraint without deeper observations in the sub-mm. An evolution of sigma could be constraints, but is not necessary to reproduce the current data.\\

The evolution of the parameters is very simple in the current version and could be updated in using more breaks or a smoother functional form.\\

The recent observations of \textit{Herschel} will help a lot to update the SED used in our model, and maybe enable to determine their evolution with the redshift. The temperature of the big grain and its dispersion will be measured more accurately. Nevertheless, this dispersion must be modeled with a limited number of template to authorize MCMC approach. It is one of the future challenge for the evolution of our model.\\

Nevertheless, each refinement increases the number of free parameters of the model. It is important to limit the number of new parameters in comparison with the number of measurements. 

\section{Summary}

\begin{itemize}
\item Our new parametric backwards evolution model reproduces the number counts from 15~$\mu$m to 1.1~mm, the monochromatic LF and the redshift distributions.
\item Our model predicts a strong evolution in luminosity of the LF up to z=2 and a strong decrease in density from z=1 to z=2. We predict that the number of HyLIRG is maximum around z=2.
\item We find that Normal galaxies, LIRG and ULIRG dominates the infrared output at z=0, z=1 and z=2, respectively. The HyLIRG accounts for a small fraction ($<$10\%) at all redshifts.
\item We reproduce the CIB spectrum and predict contributions per redshift and luminosity slice. We found that the mid- and far-infrared part of the CIB are mainly emitted by the normal galaxies and LIRG. The sub-mm part is mainly due to LIRG and ULIRG at high redshift in accordance with the sub-mm observations of deep fields. We estimated CIB total value of 23.7$\pm$0.9~nW.m$^{-2}$.sr$^{-1}$.
\item We estimate the fraction of lensed sources in the sub-mm as a function of the flux and wavelength. This contribution is low ($<10\%$) below 500~$\mu$m, but high (up to 50\%) around 100~mJy in the mm domain.
\item We predict that the population of very bright dusty galaxies detected by SPT  and without IRAS counterpart \citep{Vieira2009} is essentially composed of lensed sub-mm galaxies. We also predicts the contribution of the lensed sources to the \textit{Planck} number counts.
\item We predict the confusion limits for future missions like SPICA or CCAT.
\item We estimate the opacity of the Universe to TeV photons.
\item Material of the model (software, tables and predictions) is available at http://www.ias.u-psud.fr/irgalaxies/. 
\end{itemize}

\section{Conclusion}

We showed that it is possible to reproduce the number counts from the mid-IR to the mm domain with a rather simple parametric model minimized automatically. Nevertheless, other automatically-tuned models reproduce the counts with different redshift distributions \citep{Le_Borgne2009,Marsden2010}. It suggests that number counts only are not enough to build these models. Different observables are thus crucial to discriminate the different parametrization proposed by the model builders. These constraints are the luminosity functions, the redshift distributions, the P(D) or the fluctuations. These future measurements and their uncertainties have to be very robust to be directly fit by the next generation of models.

\begin{acknowledgements}
We acknowledge Mattia Negrello for explaining us how includes the lensing in our galaxy evolution model, Guillaume Patanchon for pushing us to make a parametric model, Nicolas Taburet and Marian Douspis for their explanations about the MCMC, Julien Grain for his explanations on the photon-photon interaction, Josh Younger for giving us useful references, Axel Wei$\beta$ and Alexandre Beelen for their discussion about the counts in the sub-mm, and Morgane Cousin for carefully reading the draft and founding some mistakes. We also thank the BLAST team for the public release of their maps. We finally acknowledge Stefano Berta, Seb Oliver, Dave Clements, Axel Wei$\beta$, Rosalind Hopwood, Joaquin Vieira, Andrew Hopkins, Alexandre Beelen and Kirsten Knudsen and Vandana Desai for providing us quickly their results. We also thanks Gaelen Marsden for our discussion about the comparison of our two models after the submission of our papers. This work was partially supported by the ANR-09-BLAN-0224-02.

\end{acknowledgements}

\bibliographystyle{aa}

\bibliography{Bethermin_model}

\begin{table}
\begin{tabular}{lrrrr}
\hline
\hline
$\lambda$ & $5\sigma_{conf,P(D)}$ & CIB fraction\tablefootmark{a} & $S_{conf,density}$ & CIB fraction\tablefootmark{b} \\
$\mu$m & mJy & \% & mJy & \% \\
\hline
24 & 5.62$\times$10$^{-2}$ &  83.1 & 7.51$\times$10$^{-2}$ &  72.3 \\ 
70 & 3.09$\times$10$^{0}$ &  51.5 & 2.88$\times$10$^{0}$ &  48.8 \\ 
100 & 1.38$\times$10$^{1}$ &  36.3 & 1.15$\times$10$^{1}$ &  36.1 \\ 
160 & 5.84$\times$10$^{1}$ &  12.3 & 3.43$\times$10$^{1}$ &  17.2 \\ 
250 & 1.06$\times$10$^{2}$ &   3.2 & 4.41$\times$10$^{1}$ &   6.9 \\ 
350 & 1.13$\times$10$^{2}$ &   0.8 & 3.57$\times$10$^{1}$ &   3.0 \\ 
500 & 9.18$\times$10$^{1}$ &   0.2 & 2.24$\times$10$^{1}$ &   1.4 \\ 
850 & 4.12$\times$10$^{1}$ & 100.0 & 9.25$\times$10$^{0}$ &   0.7 \\ 
1100 & 2.76$\times$10$^{1}$ & 100.0 & 6.25$\times$10$^{0}$ &   0.5 \\ 
\hline
\end{tabular}
\tablefoot{
\tablefoottext{a}{Fraction of the CIB resolved at 5-$\sigma_{conf}$}. \\
\tablefoottext{b}{Fraction of the CIB resolved at the flux limit.   }
}
\caption{\label{tab:0085} Confusion noise and resolved fraction of the CIB at different wavelengths for a  0.85~m telescope(\textit{Spitzer} like).}
\end{table}

\begin{table}
\begin{tabular}{lrrrr}
\hline
\hline
$\lambda$ & $5\sigma_{conf,P(D)}$ & CIB fraction\tablefootmark{a} & $S_{conf,density}$ & CIB fraction\tablefootmark{b} \\
$\mu$m & mJy & \% & mJy & \% \\
\hline
70 & 7.95$\times$10$^{-2}$ &  96.4 & 1.20$\times$10$^{-1}$ &  91.8 \\ 
100 & 5.13$\times$10$^{-1}$ &  90.8 & 7.75$\times$10$^{-1}$ &  83.9 \\ 
160 & 5.01$\times$10$^{0}$ &  67.8 & 5.93$\times$10$^{0}$ &  59.8 \\ 
250 & 1.75$\times$10$^{1}$ &  25.9 & 1.28$\times$10$^{1}$ &  29.6 \\ 
350 & 2.30$\times$10$^{1}$ &   9.2 & 1.28$\times$10$^{1}$ &  15.8 \\ 
500 & 2.08$\times$10$^{1}$ &   3.3 & 9.24$\times$10$^{0}$ &   8.7 \\ 
850 & 1.13$\times$10$^{1}$ &   1.5 & 3.88$\times$10$^{0}$ &   4.4 \\ 
1100 & 8.40$\times$10$^{0}$ &   1.2 & 2.66$\times$10$^{0}$ &   3.5 \\ 
\hline
\end{tabular}
\tablefoot{
\tablefoottext{a}{Fraction of the CIB resolved at 5-$\sigma_{conf}$}. \\
\tablefoottext{b}{Fraction of the CIB resolved at the flux limit.   }
}
\caption{\label{tab:0329} Confusion noise and resolved fraction of the CIB at different wavelengths for a  3.29~m telescope (\textit{Herschel} like).}
\end{table}

\begin{table}
\begin{tabular}{lrrrr}
\hline
\hline
$\lambda$ & $5\sigma_{conf,P(D)}$ & CIB fraction\tablefootmark{a} & $S_{conf,density}$ & CIB fraction\tablefootmark{b} \\
$\mu$m & mJy & \% & mJy & \% \\
\hline
160 & 5.86$\times$10$^{-2}$ &  99.4 & 5.55$\times$10$^{-2}$ &  98.2 \\ 
250 & 7.06$\times$10$^{-1}$ &  94.2 & 1.11$\times$10$^{0}$ &  85.6 \\ 
350 & 2.08$\times$10$^{0}$ &  77.9 & 2.57$\times$10$^{0}$ &  63.2 \\ 
500 & 3.05$\times$10$^{0}$ &  50.0 & 2.57$\times$10$^{0}$ &  41.8 \\ 
850 & 2.19$\times$10$^{0}$ &  23.6 & 1.24$\times$10$^{0}$ &  22.9 \\ 
1100 & 1.74$\times$10$^{0}$ &  18.4 & 8.74$\times$10$^{-1}$ &  18.6 \\ 
\hline
\end{tabular}
\tablefoot{
\tablefoottext{a}{Fraction of the CIB resolved at 5-$\sigma_{conf}$}. \\
\tablefoottext{b}{Fraction of the CIB resolved at the flux limit.   }
}
\caption{\label{tab:1200} Confusion noise and resolved fraction of the CIB at different wavelengths for a 12.00~m telescope (APEX like).}
\end{table}

\begin{table}
\begin{tabular}{lrrrr}
\hline
\hline
$\lambda$ & $5\sigma_{conf,P(D)}$ & CIB fraction\tablefootmark{a} & $S_{conf,density}$ & CIB fraction\tablefootmark{b} \\
$\mu$m & mJy & \% & mJy & \% \\
\hline
160 & 2.34$\times$10$^{-2}$ &  99.8 & 1.04$\times$10$^{-2}$ &  99.3 \\ 
250 & 3.01$\times$10$^{-1}$ &  97.6 & 4.48$\times$10$^{-1}$ &  92.5 \\ 
350 & 1.08$\times$10$^{0}$ &  88.6 & 1.55$\times$10$^{0}$ &  74.7 \\ 
500 & 1.87$\times$10$^{0}$ &  66.6 & 1.86$\times$10$^{0}$ &  52.4 \\ 
850 & 1.55$\times$10$^{0}$ &  33.8 & 9.70$\times$10$^{-1}$ &  29.4 \\ 
1100 & 1.26$\times$10$^{0}$ &  26.7 & 6.89$\times$10$^{-1}$ &  24.1 \\ 
\hline
\end{tabular}
\tablefoot{
\tablefoottext{a}{Fraction of the CIB resolved at 5-$\sigma_{conf}$}. \\
\tablefoottext{b}{Fraction of the CIB resolved at the flux limit.   }
}
\caption{\label{tab:1500} Confusion noise and resolved fraction of the CIB at different wavelengths for a 15.00~m telescope (CSO like).}
\end{table}

\begin{table}
\begin{tabular}{lrrrr}
\hline
\hline
$\lambda$ & $5\sigma_{conf,P(D)}$ & CIB fraction\tablefootmark{a} & $S_{conf,density}$ & CIB fraction\tablefootmark{b} \\
$\mu$m & mJy & \% & mJy & \% \\
\hline
250 & 2.81$\times$10$^{-2}$ &  99.8 & 1.32$\times$10$^{-2}$ &  99.1 \\ 
350 & 1.57$\times$10$^{-1}$ &  98.5 & 2.12$\times$10$^{-1}$ &  94.2 \\ 
500 & 4.31$\times$10$^{-1}$ &  92.6 & 6.09$\times$10$^{-1}$ &  79.1 \\ 
850 & 5.99$\times$10$^{-1}$ &  64.6 & 4.62$\times$10$^{-1}$ &  49.7 \\ 
1100 & 5.39$\times$10$^{-1}$ &  53.1 & 3.46$\times$10$^{-1}$ &  41.2 \\ 
\hline
\end{tabular}
\tablefoot{
\tablefoottext{a}{Fraction of the CIB resolved at 5-$\sigma_{conf}$}. \\
\tablefoottext{b}{Fraction of the CIB resolved at the flux limit.   }
}
\caption{\label{tab:2500} Confusion noise and resolved fraction of the CIB at different wavelengths for a 25.00~m telescope (CCAT like).}
\end{table}

\end{document}